\documentclass{emulateapj}
\usepackage{ulem}
\usepackage{longtable}
\usepackage{graphicx}
\usepackage{txfonts}
\usepackage{colortbl}
\usepackage{multirow}
\usepackage{bm}

\begin{document}
\title{Modeling gamma-ray light curves with more realistic pulsar magnetospheres }

\author{Gang Cao\altaffilmark{1} and Xiongbang Yang\altaffilmark{2}}
\altaffiltext{1}{Department of  Mathematics, Yunnan University of Finance and Economics, Kunming 650221, Yunnan, P. R. China}
\altaffiltext{2}{Department of Astronomy, Yunnan University, Key Laboratory of Astroparticle Physics of Yunnan Province, Kunming 650091, Yunnan, P. R. China}
\email{E-mail: gcao@ynao.ac.cn}
\shortauthors{Cao et al.}
\shorttitle{Modeling gamma-ray light curves}

\begin{abstract}
We study the gamma-ray emission patterns and light curves in dissipative pulsar magnetospheres.
We produce the gamma-ray light curves by using the geometric method and the particle trajectory method. For the geometric method, assuming the gamma-ray emission originates in a finite-width layer  along the last closed lines, we generate the gamma-ray light curves based on the uniform emissivity along the magnetic field lines in the comoving frame (CF). For the particle trajectory method, we consider the spatial distribution of conductivity $\sigma$ by assuming a very high conductivity within the light cylinder (LC) and a finite conductivity outside the LC . Assuming that all the $\gamma$-ray emission originates in the outer magnetosphere outside the LC, we generate the gamma-ray light curves by computing realistic particle trajectories and Lorentz factors, taking into account both the accelerating electric field and curvature radiation loss. Further, we compare the modeling light curves to the observed light curves at $>0.1\, \rm GeV$ energies for Vela pulsar. Our results show that the magnetosphere with the low $\sigma$ value is more preferred for the geometric method. However, the magnetosphere with a near force-free regime within the LC and a high $\sigma$ value outside the LC is more favored for the particle trajectory method. It is noted that the particle trajectory method uses the parallel electric fields that are self-consistent with the magnetic fields of the magnetosphere. We suggest that the results from the particle trajectory method are more supported on the physical ground.
\end{abstract}


\section{Introduction}
The launch of Fermi Gamma-Ray Space Telescope has opened a new era in the study of pulsar physics. In the first year of operation of the Large Area Telescope (LAT), more than 40 new gamma-ray pulsars were discovered \citep{abd10}. To date, more than 150 gamma-ray pulsars have been detected by the LAT \citep{ack15}, 117 of which are included in the Second Fermi Pulsar Catalog \citep{abd13}. They are divided into three groups: millisecond pulsars, young radio-loud pulsars and young radio-quiet pulsars. High-quality light curves, phase-averaged spectra and phase-resolved spectra obtained by the LAT observations provide an opportunity of understanding the emission sites and magnetic field geometries of these gamma-ray pulsars.

The realistic structures of the pulsar magnetosphere still remain uncertain. Knowledge about the pulsar magnetosphere structures can be used to identify the potential sites of particle acceleration and gamma-ray emission. A vacuum dipole field is generally adopted in the early study of pulsar emission, because it has an exact analytical solution given by \citet{deu55}. Based on this field structure, different theoretical models have been developed to explain the observed pulsar emission. In these models, it is widely believed that particles are accelerated in the gap region where an accelerating electric field is created because of the deficit of charges. Gamma-rays emission is produced by curvature or inverse-Compton radiation from high-energy particles accelerated in these gaps. Due to different emission zone locations, standard pulsar radiation models include the polar cap \citep[e.g., ][]{rud75,dau82}, the slot-gap (SG) \citep[e.g., ][]{dyk03,dyk04,mus04}, and the outer-gap (OG) \citep[e.g., ][]{che86,zha97,che00,zha01,zha04} models. These gap models have achieved great successes in explaining pulsar high-energy emissions and light curves \citep[e.g., ][]{wat09,rom10}. However, the vacuum solution has no plasma, it is not able to reproduce any pulsar phenomenons. It is well known that the pulsar magnetosphere should be filled with plasma \citep{gol69}. In the presence of abundant plasma, all accelerating electric fields can be efficiently screened to form a force-free (FF) magnetosphere.
The FF solution for an aligned rotator was first obtained by \citet{con99}. The CKF solution consists of a closed field line region extending to the LC, an open field line region, and an equatorial current sheet beyond the LC. Moreover, the time-dependent simulations for the FF axisymmetric rotator also confirmed the closed-open CKF solution \citep[e.g.,][]{kom06,mck06,tim06,yu11,par12,cao16a,eti16}.

\begin{figure*}
\begin{tabular}{cccccc}
  \includegraphics[width=5.5cm,height=5.5cm]{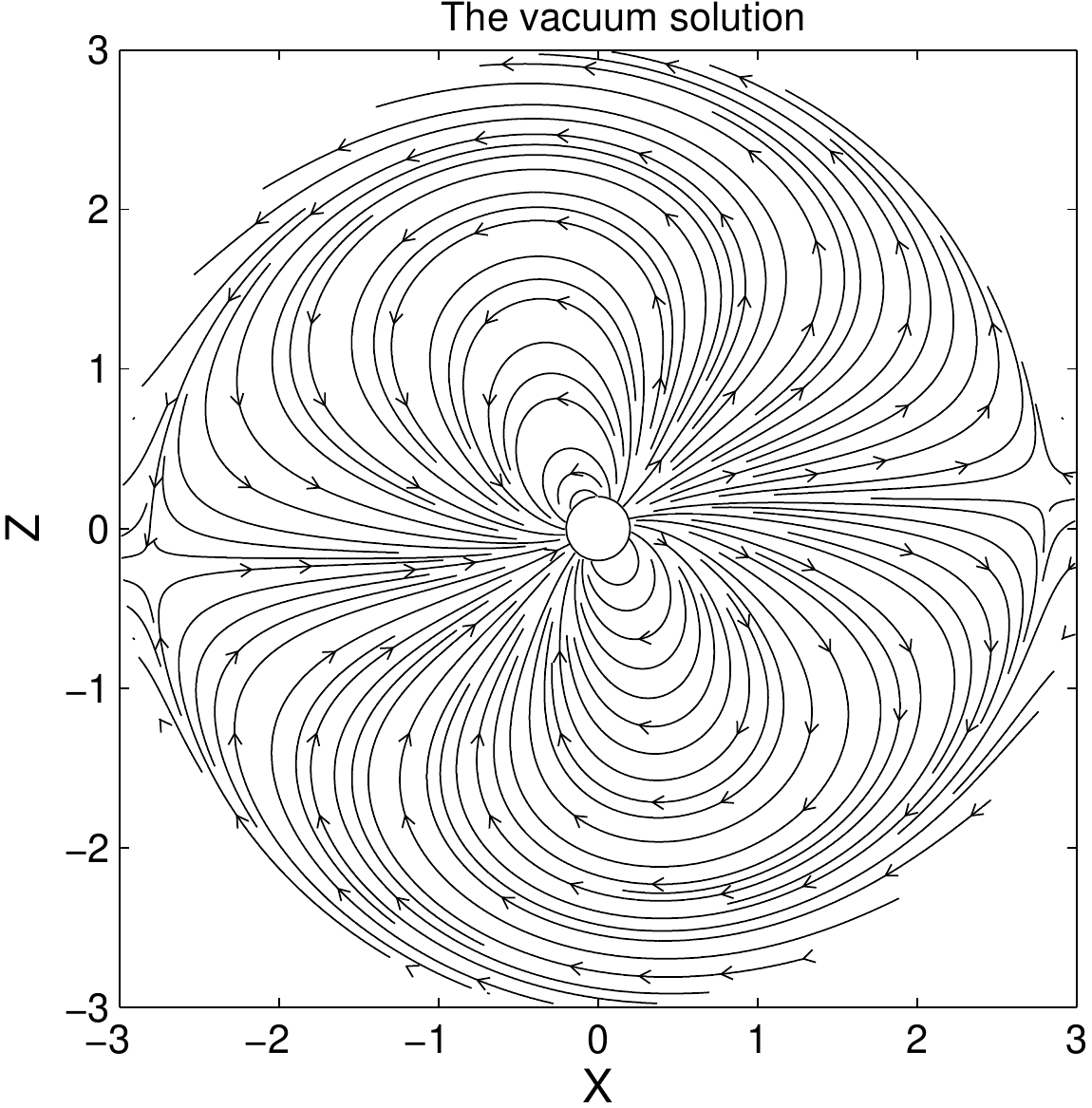} \qquad
  \includegraphics[width=5.5cm,height=5.5cm]{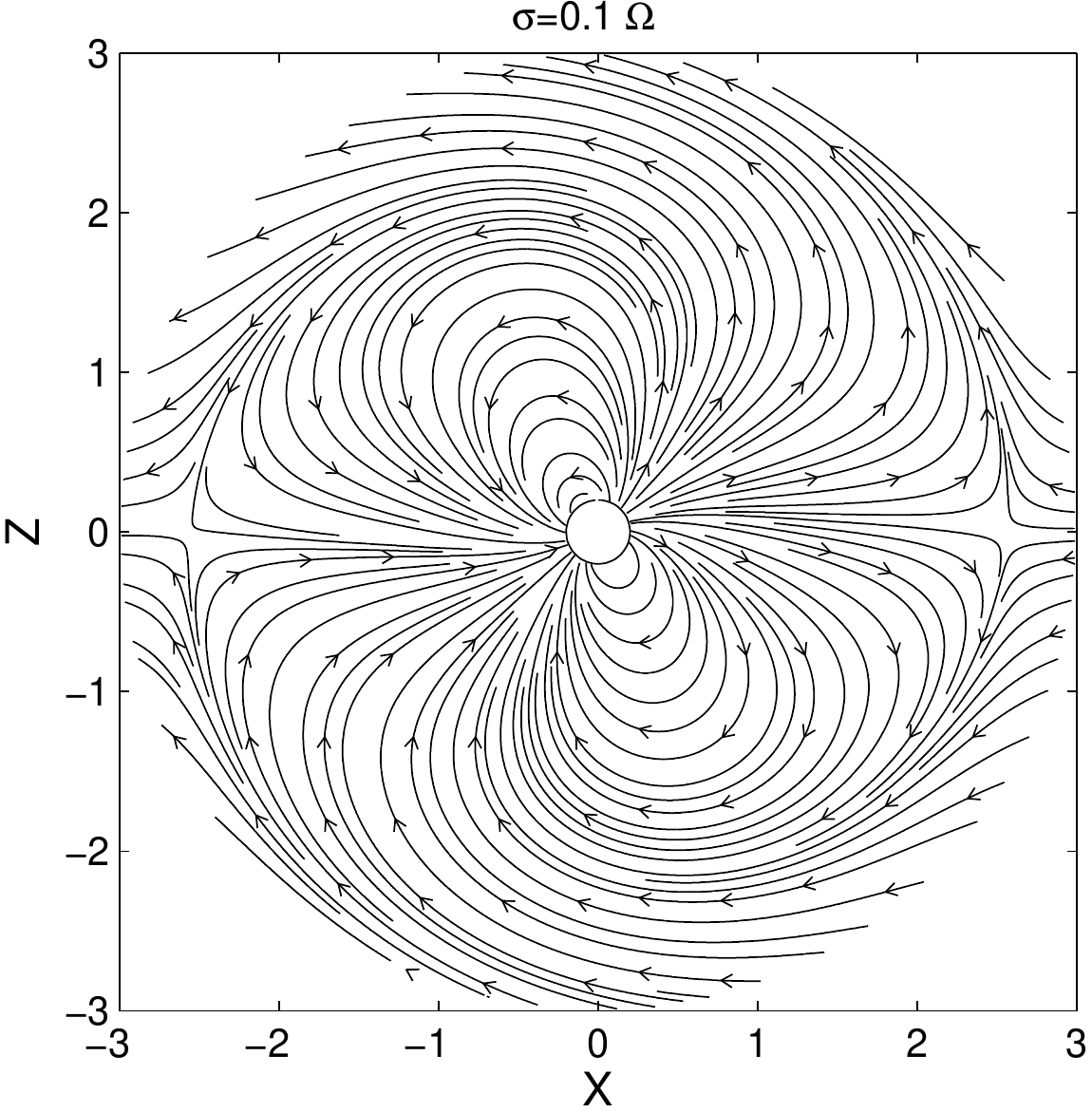} \qquad
  \includegraphics[width=5.5cm,height=5.5cm]{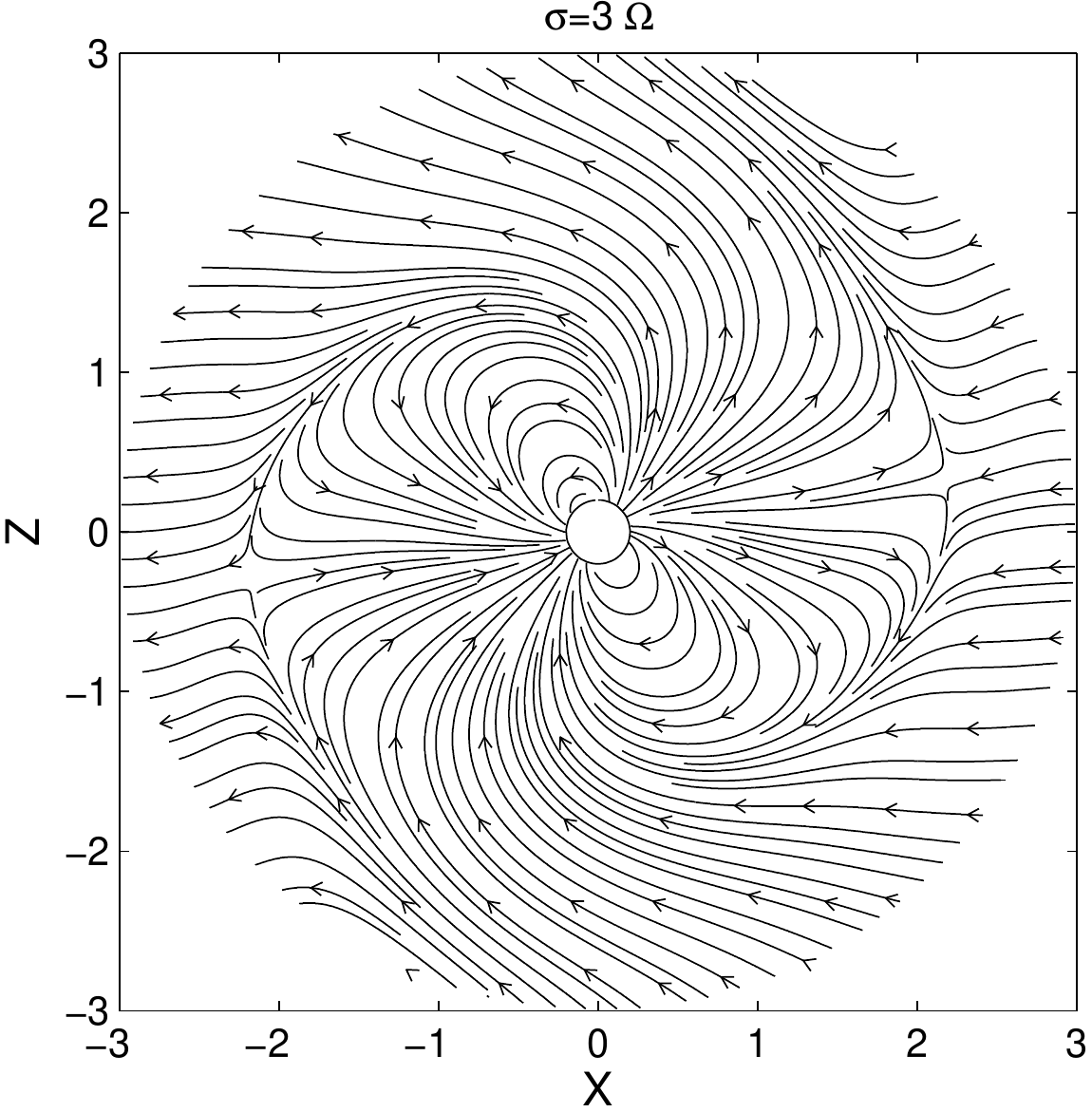} \\\\
  \includegraphics[width=5.5cm,height=5.5cm]{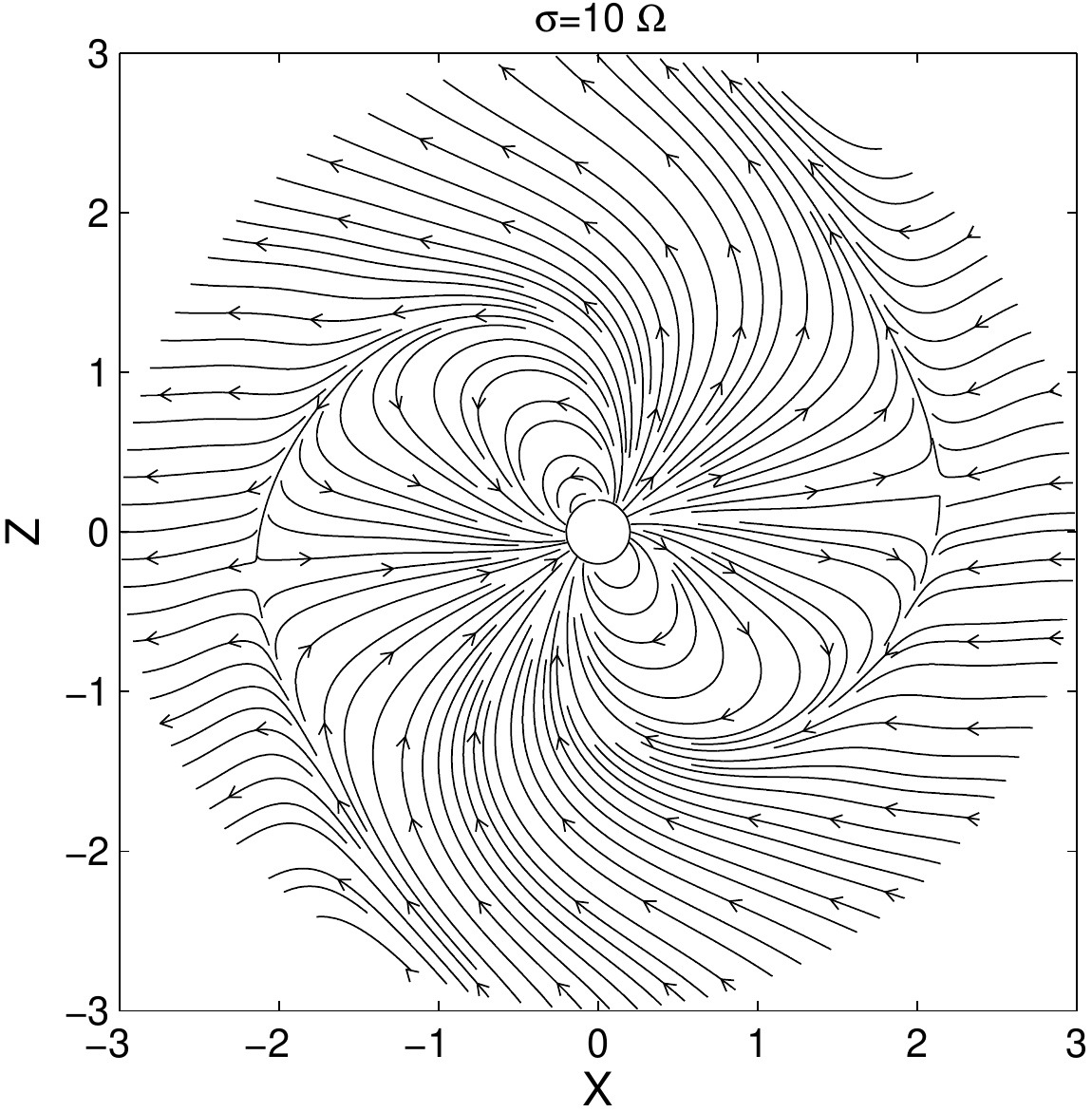} \qquad
  \includegraphics[width=5.5cm,height=5.5cm]{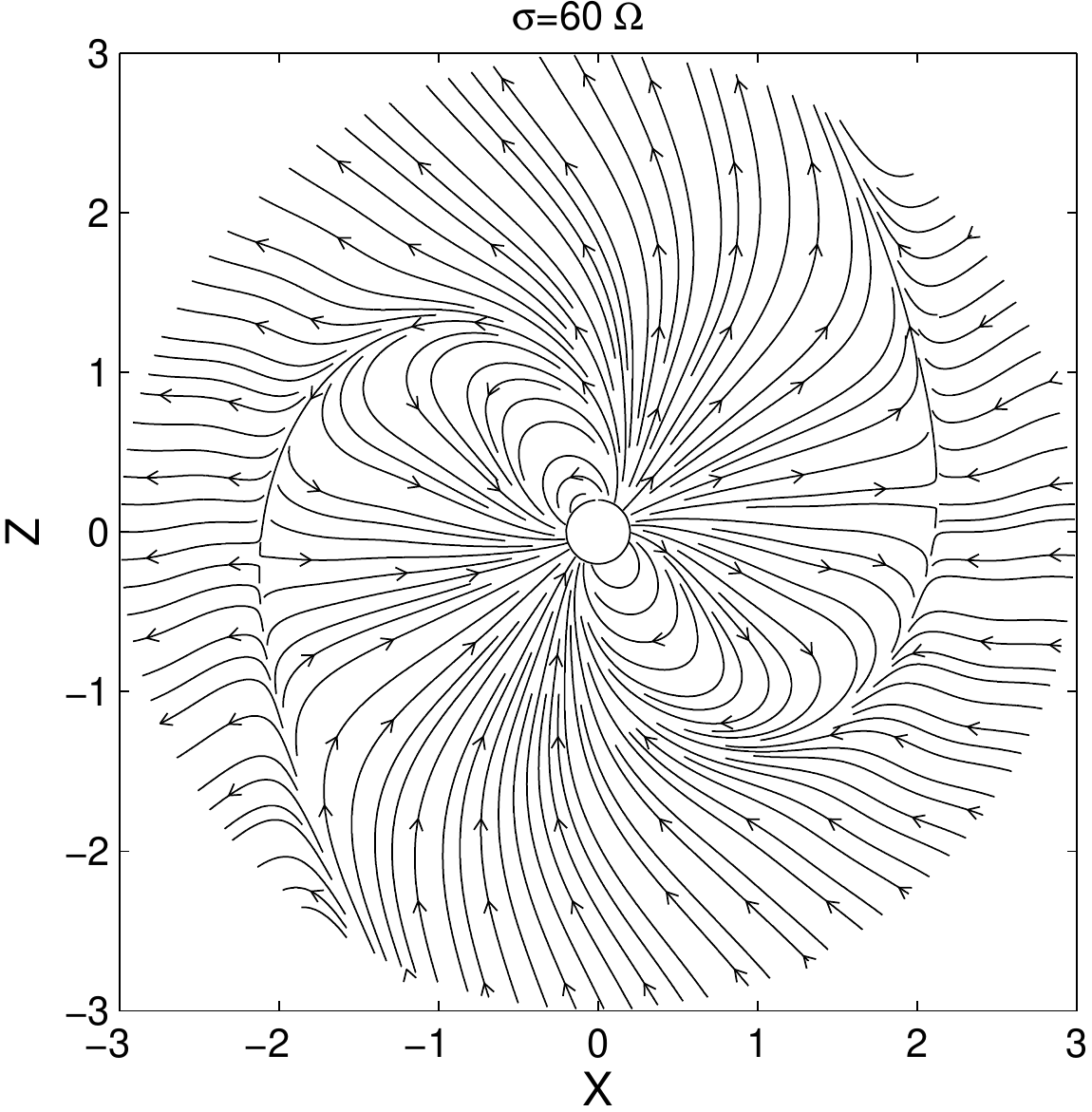} \qquad
  \includegraphics[width=5.5cm,height=5.5cm]{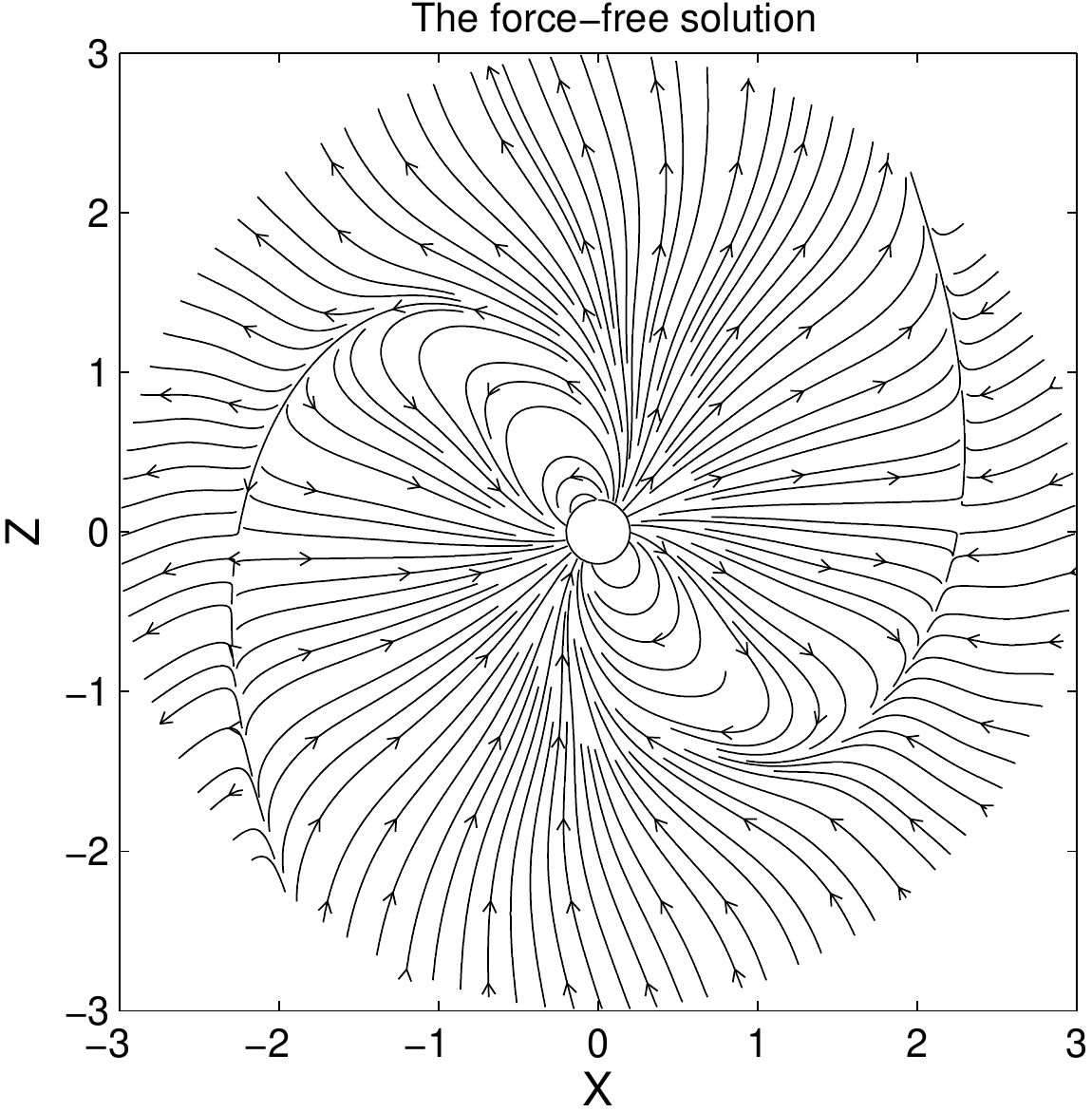} \\
\end{tabular}
\caption{Magnetic field structures in the $\Omega-\mu$ plane for magnetic inclination $\alpha=60^{\circ}$ with increasing conductivity $\sigma$ from $\sigma=0$ to $\sigma \rightarrow \infty$ }
\end{figure*}

\begin{figure}
\center
\begin{tabular}{cccccc}
\includegraphics[width=7.5cm,height=6.6cm]{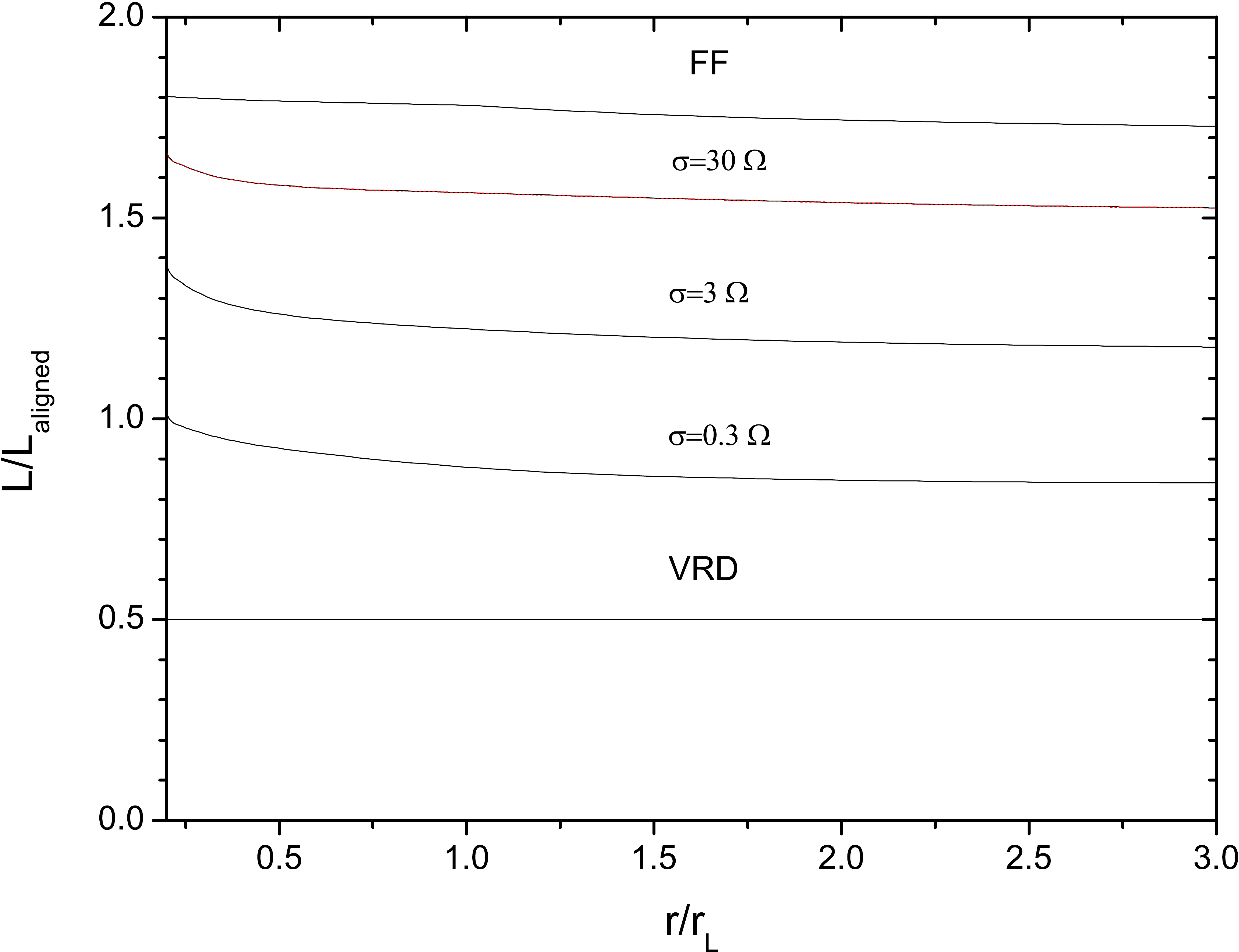}
\end{tabular}
\caption{The normalized Poynting flux $L/L_{\rm aligned}$ as a function of radius $r$ for magnetic inclination $\alpha=60^{\circ}$ with different $\sigma$ values at time $t= 3 \, P$. The red dashed curve represents the normalized Poynting flux for $\sigma=30 \, \Omega$ at time $t= 4 \, P$. }
\end{figure}

\begin{figure}
\center
\begin{tabular}{cccccc}
\includegraphics[width=7.5cm,height=6.65cm]{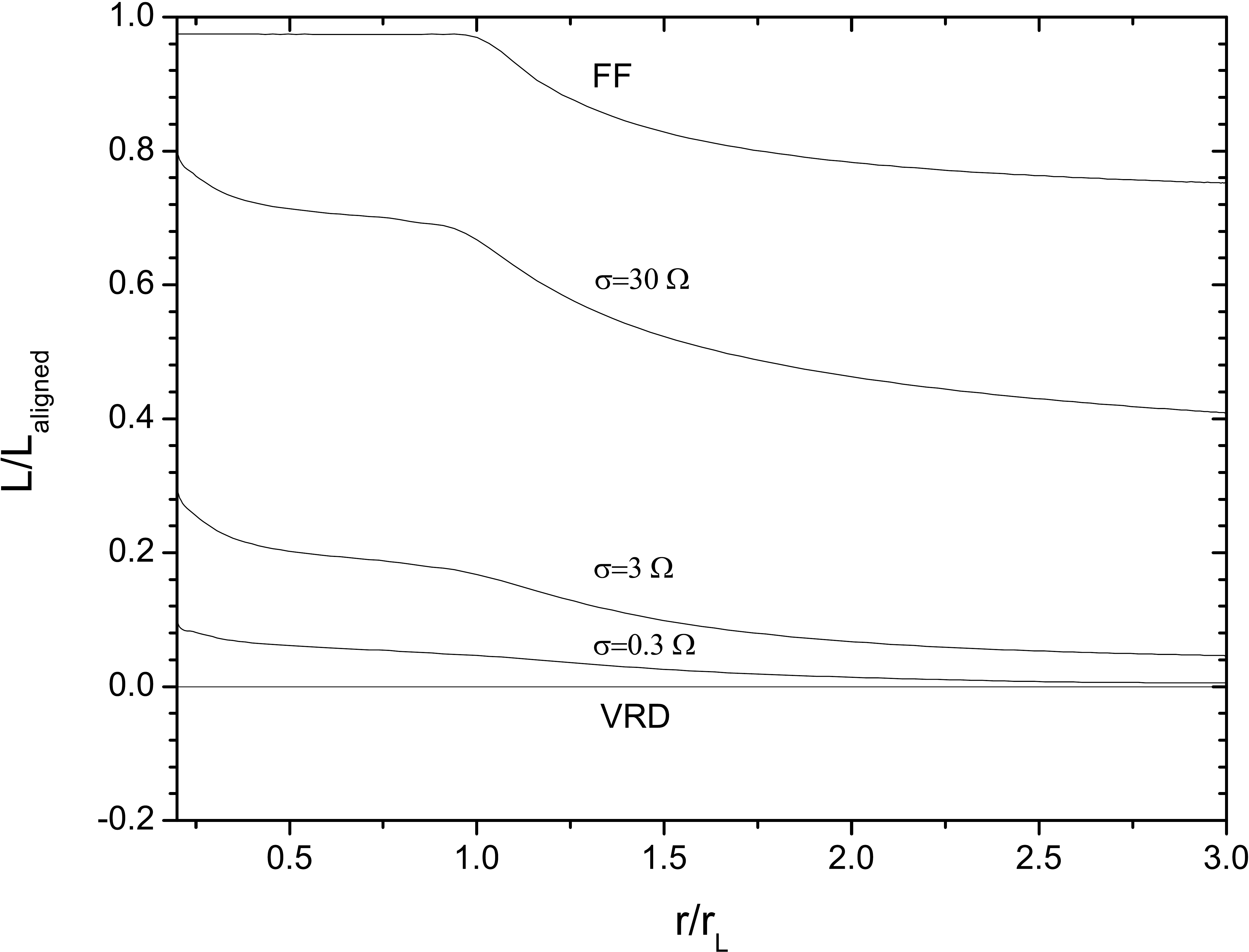}
\end{tabular}
\caption{The normalized Poynting flux $L/L_{\rm aligned}$ as a function of radius $r$ for magnetic inclination $\alpha=0^{\circ}$ with different $\sigma$ values at time $t= 3 \, P$. }
\end{figure}
Three-dimensional (3D) structures of the force-free pulsar magnetosphere have been available in recent years \citep[e.g.,][]{spi06,kal09,pet12}. Moreover, these studies have
been also extended to the general-relativistic force-free regime that takes time-space curvature and frame-dragging effects into accounts \citep{pet16,car18} and the full magnetohydrodynamic (MHD) regime that takes plasma pressures and inertial into account\citep{tch13}.
The 3D force-free solutions are similar to the CKF solution with an equatorial current sheet outside the LC. These simulations gave an impetus to the study of pulsar high-energy radiation with a more realistic field geometry instead of the vacuum dipole field. Gamma-ray light curves have been modeled using the force-free magnetic field \citep{bai10b,bog18}.
The force-free solutions provide different pulse profiles due to the increased polar cap size compared to that of the vacuum dipole. It should be noted that the force-free solutions cannot allow any accelerating electric fields. Therefore, they cannot provide any information about the sites of particle acceleration and radiation.

It is well known that the vacuum solution has no any particle distributions, while the force-free solution does not allow particle acceleration along magnetic field lines. Therefore, more realistic pulsar magnetosphere should lie between the vacuum and force-free limits. In fact, the plasma resistivity can produce a none-zero accelerating electric field along the magnetic field, and a sets of resistive solutions that smoothly bridges the gap between the vacuum and force-free solutions have been constructed based on the finite-difference time-domain approach \citep{li12,kal12a}. Moreover, the resistive pulsar magnetospheres have been used to model the pulsar $\gamma$-ray spectra and light curves \citep{kal14,bra15}. Also, particle-in-cell (PIC) methods with a self-consistent treatment between particles and fields are developed to simulate the structures of the pulsar magnetosphere \citep{phi14,che14,bel15,cer15,phi15,kal18,bra18}. Very recently, full PIC simulations also start to predict the pulsar light curves by including the radiation reaction \citep{cer16,phi18,kal18}.

\begin{figure}
\epsscale{1.}
\plotone{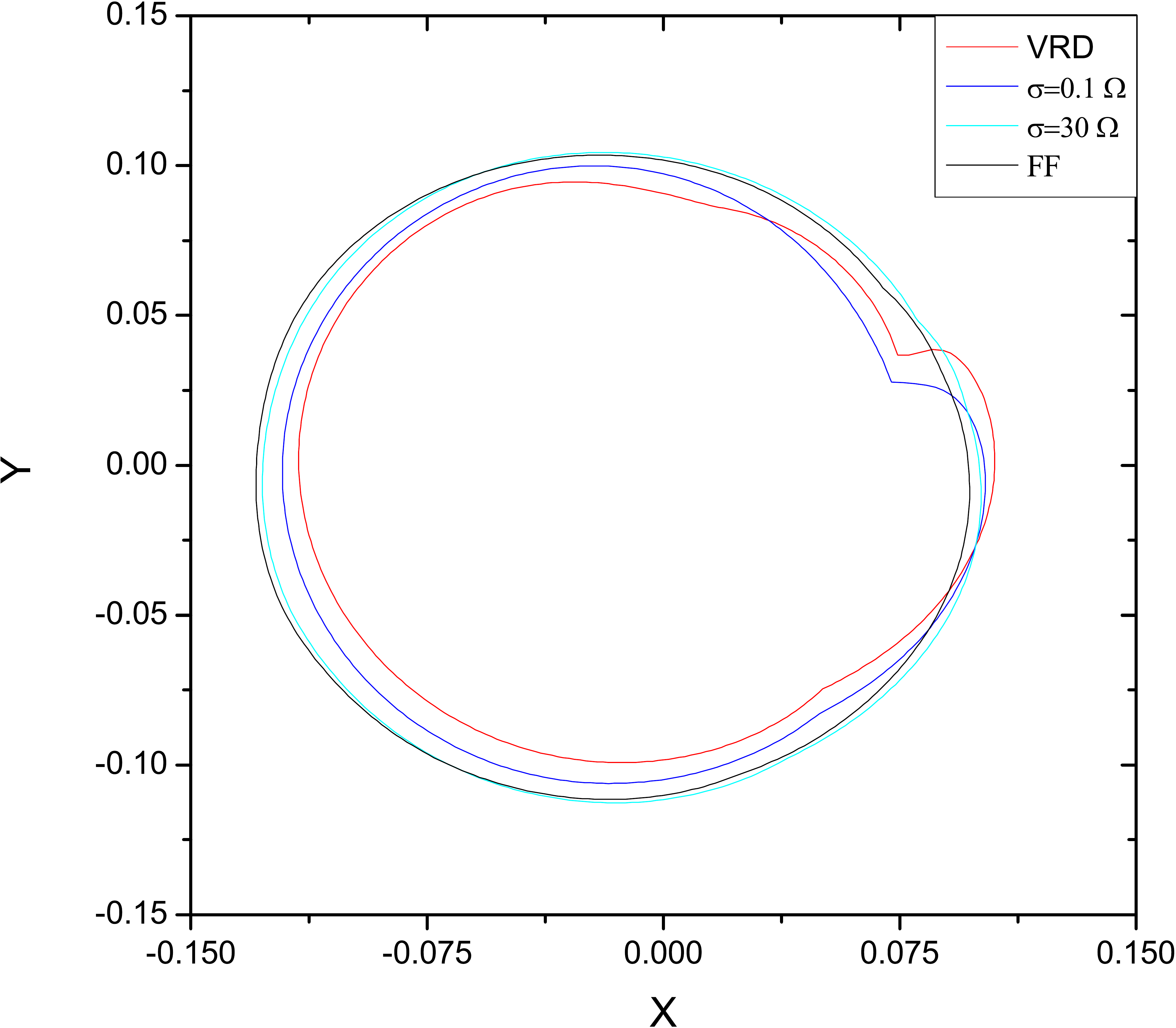}
\caption{The polar cap shapes on a sphere of radius $0.2 \, r_{\rm L}$ for magnetic inclination $\alpha=60^{\circ}$ with different $\sigma$ values. }
\end{figure}

\begin{figure*}
\begin{tabular}{cccccccc}
  \includegraphics[width=3.75 cm,height=3.75 cm]{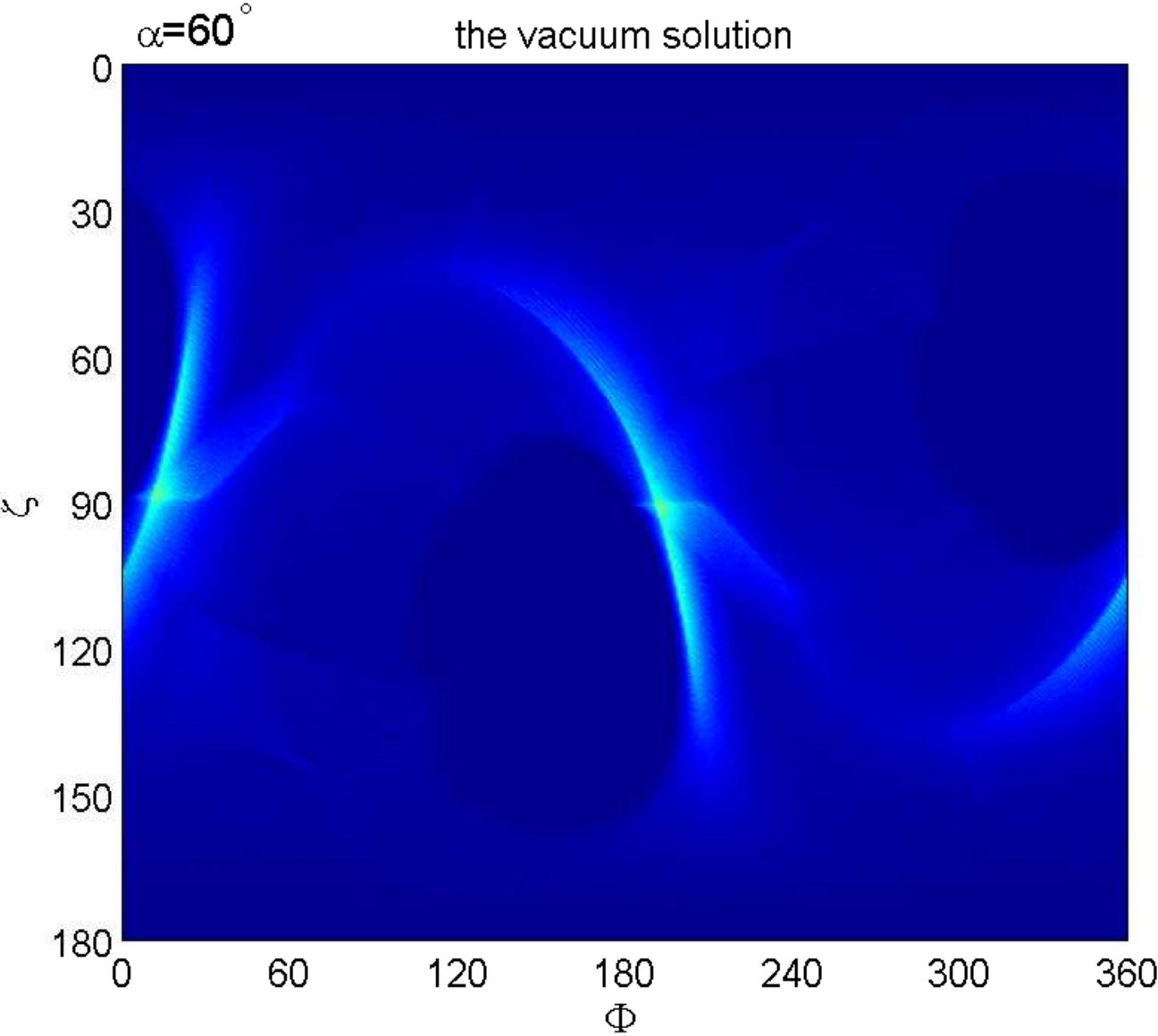} \quad \qquad
  \includegraphics[width=3.5 cm,height=3.75 cm]{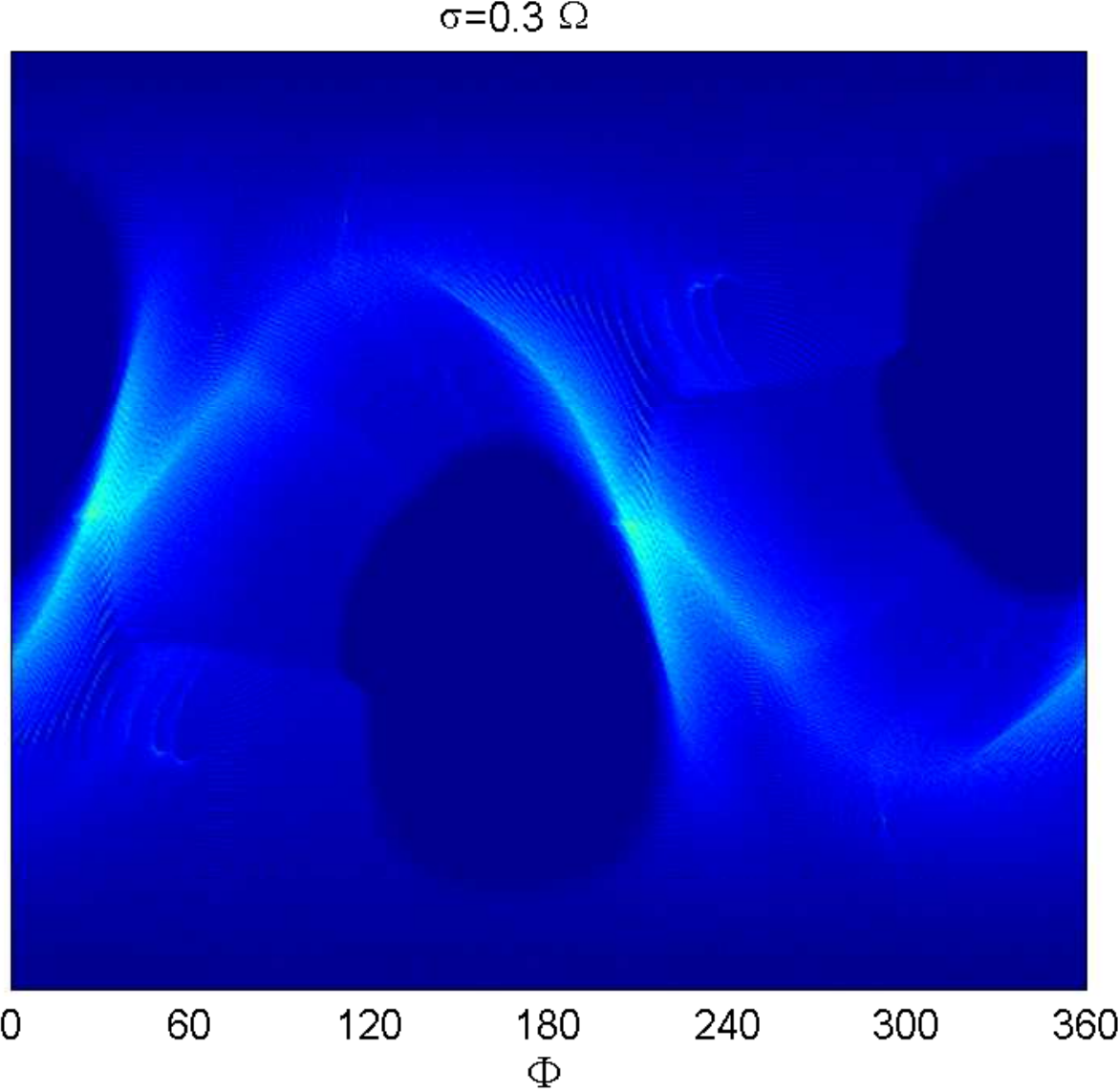} \quad \qquad
  \includegraphics[width=3.5 cm,height=3.75 cm]{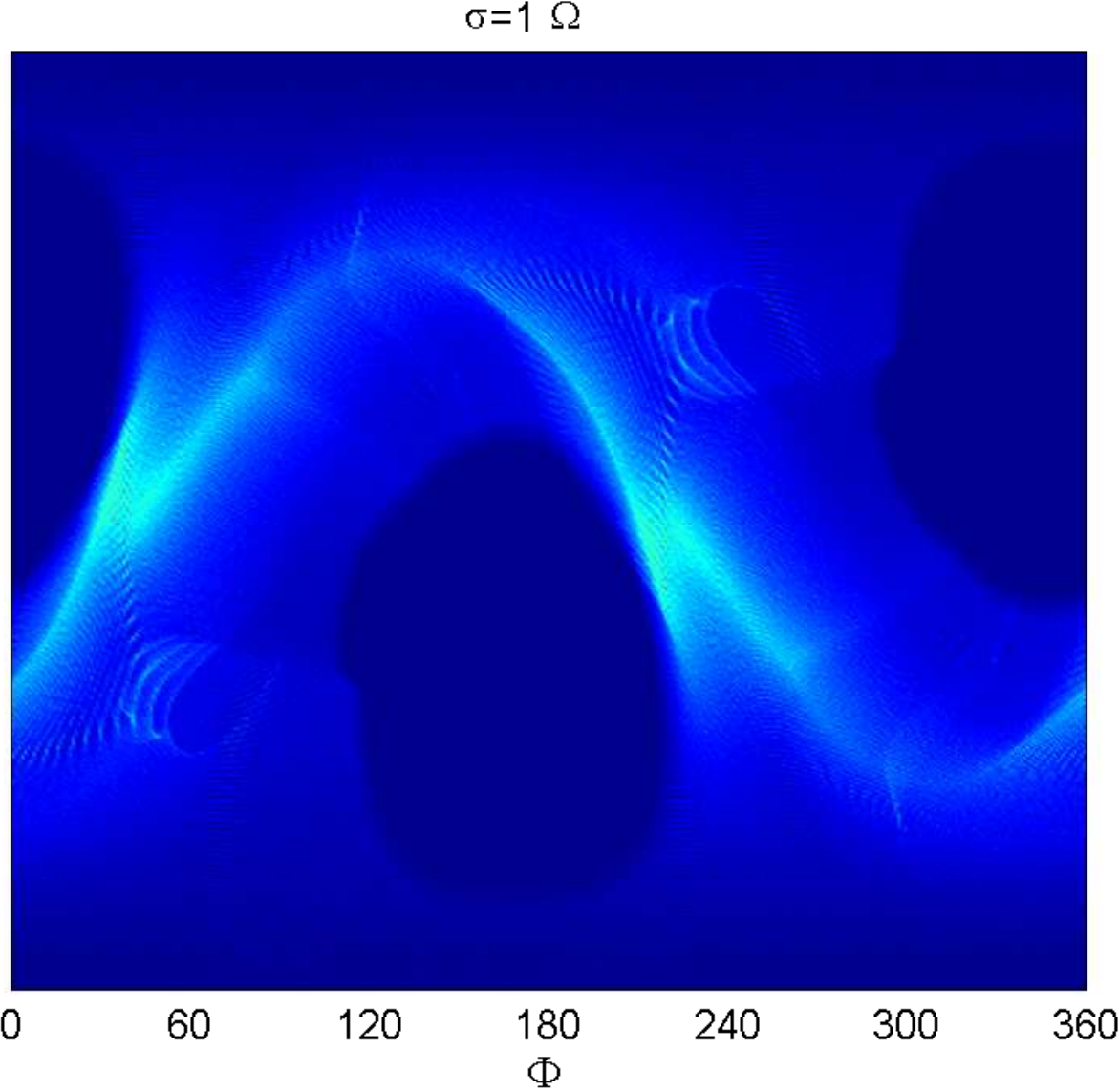} \quad \qquad
  \includegraphics[width=3.5 cm,height=3.75 cm]{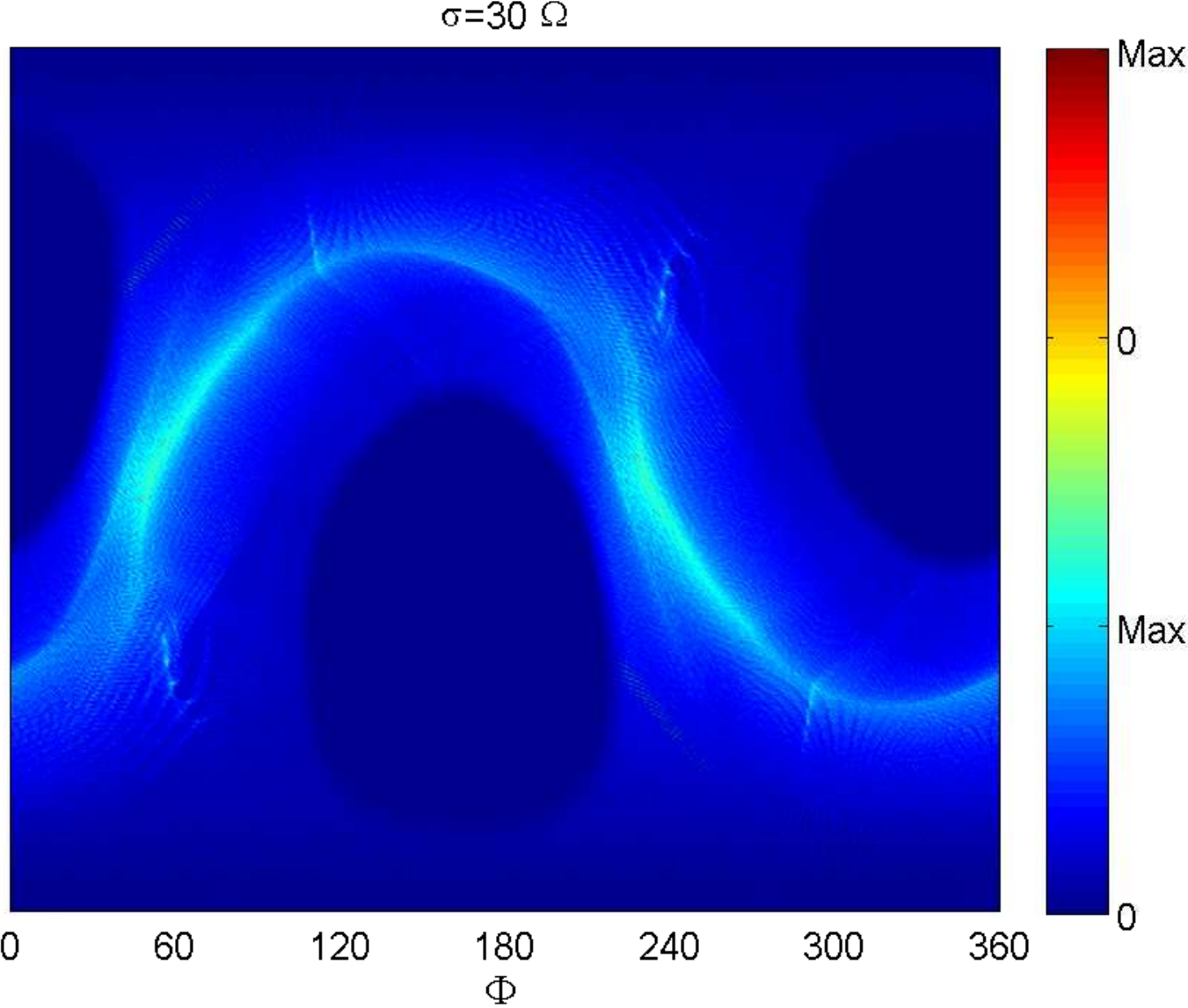} \\\\
  \includegraphics[width=3.75 cm,height=3.8 cm]{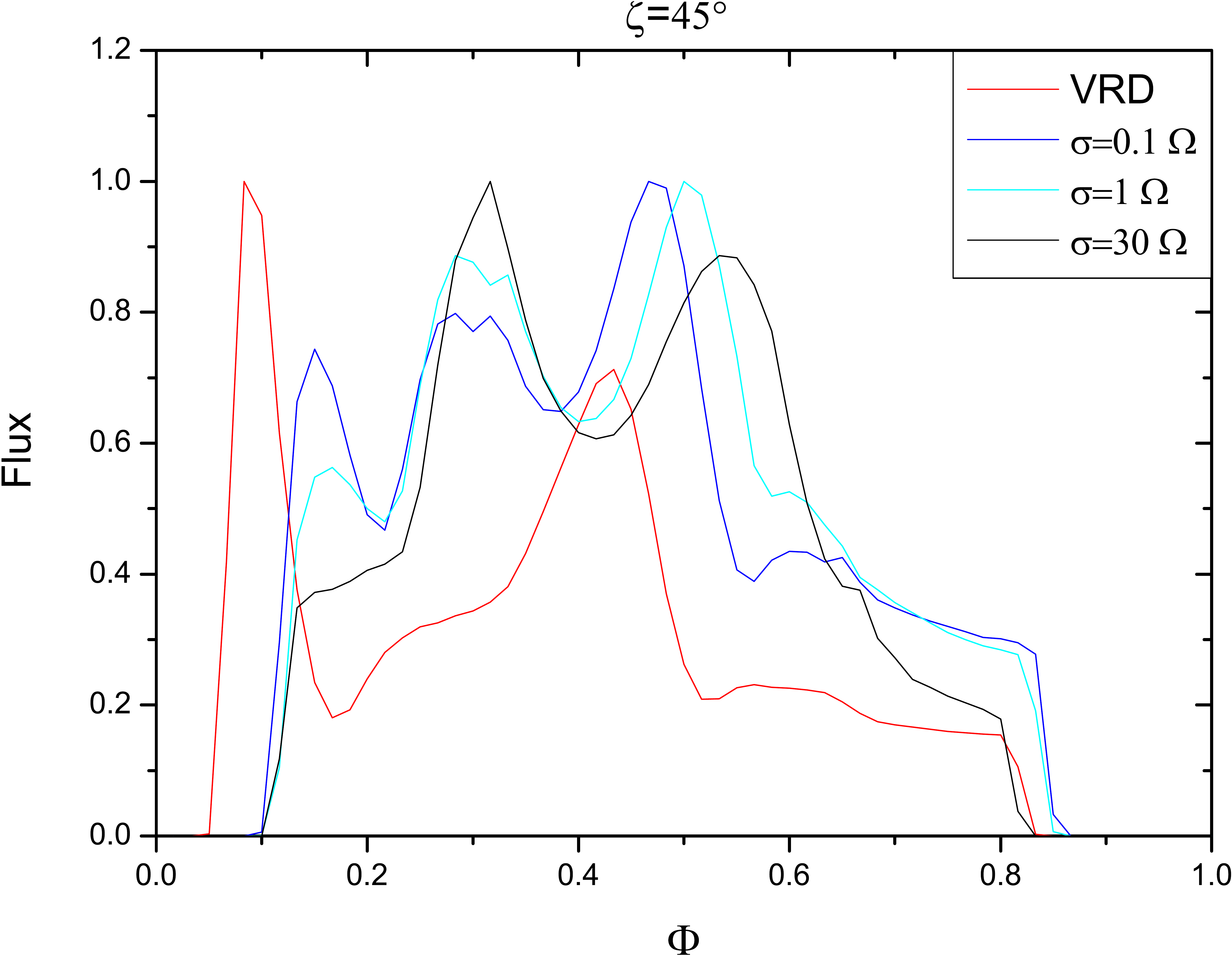} \quad \qquad
  \includegraphics[width=3.5 cm,height=3.8 cm]{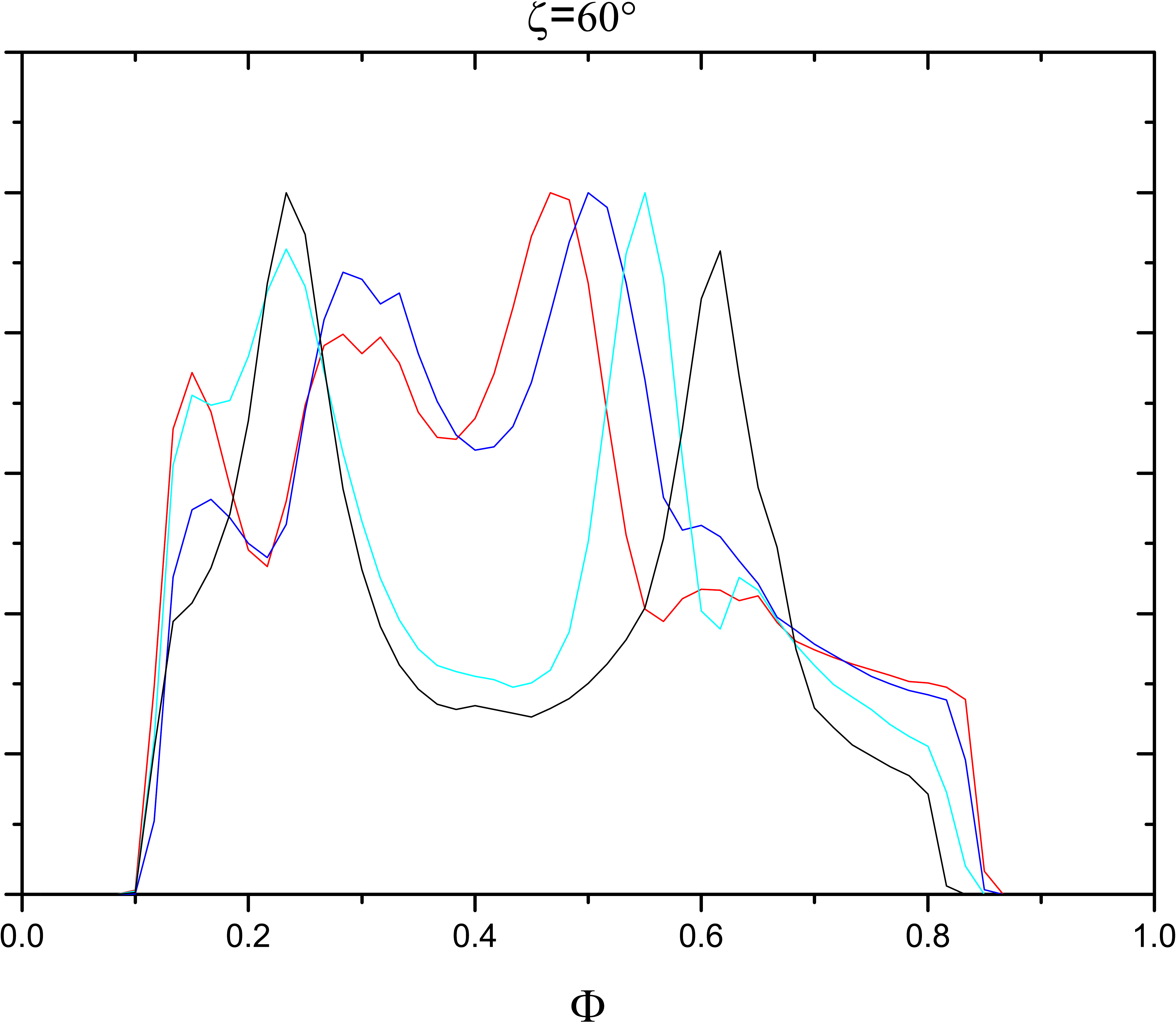} \quad \qquad
  \includegraphics[width=3.5 cm,height=3.8 cm]{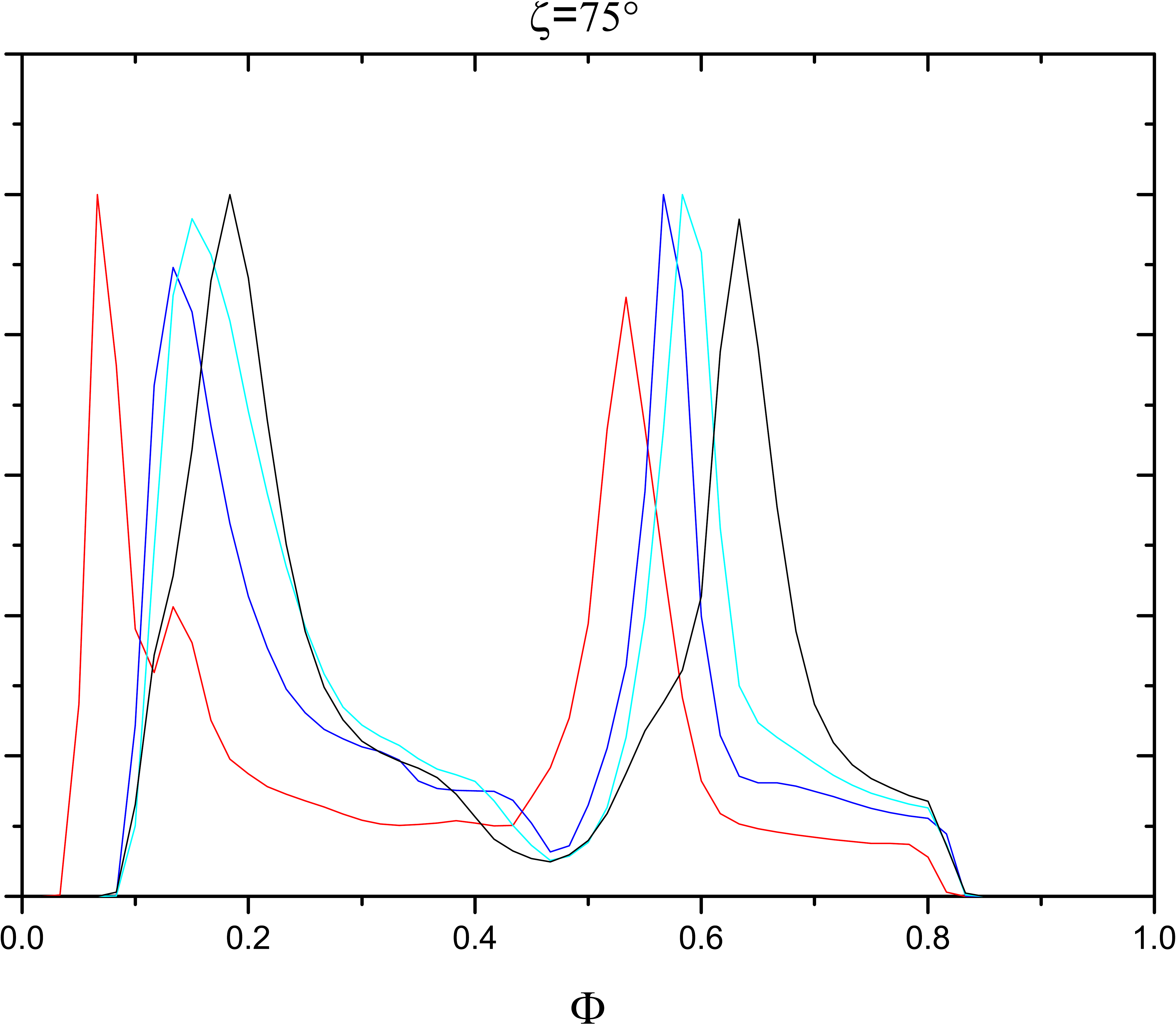} \quad \qquad
  \includegraphics[width=3.5 cm,height=3.8 cm]{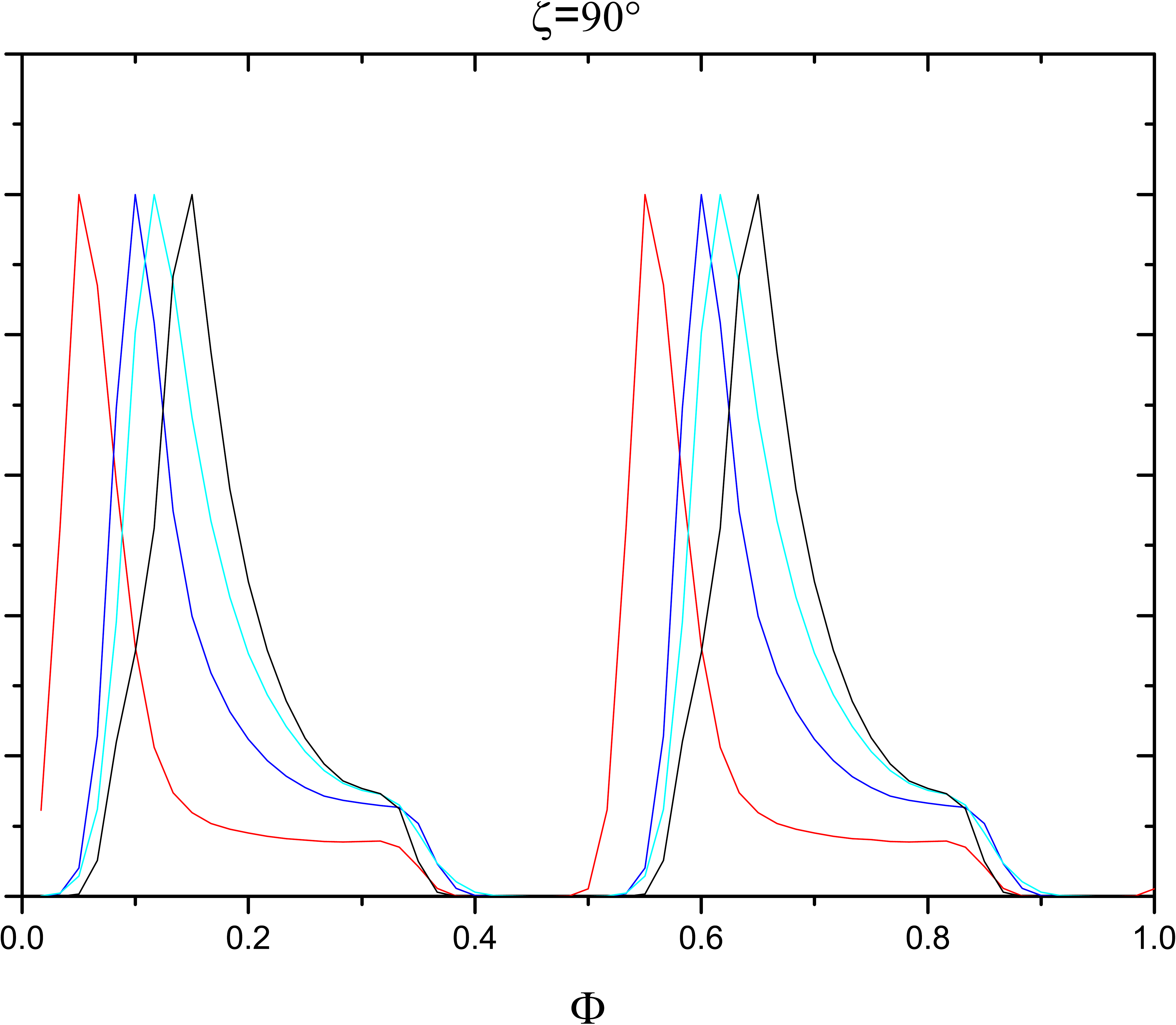} \\\\
  \includegraphics[width=3.75 cm,height=3.75 cm]{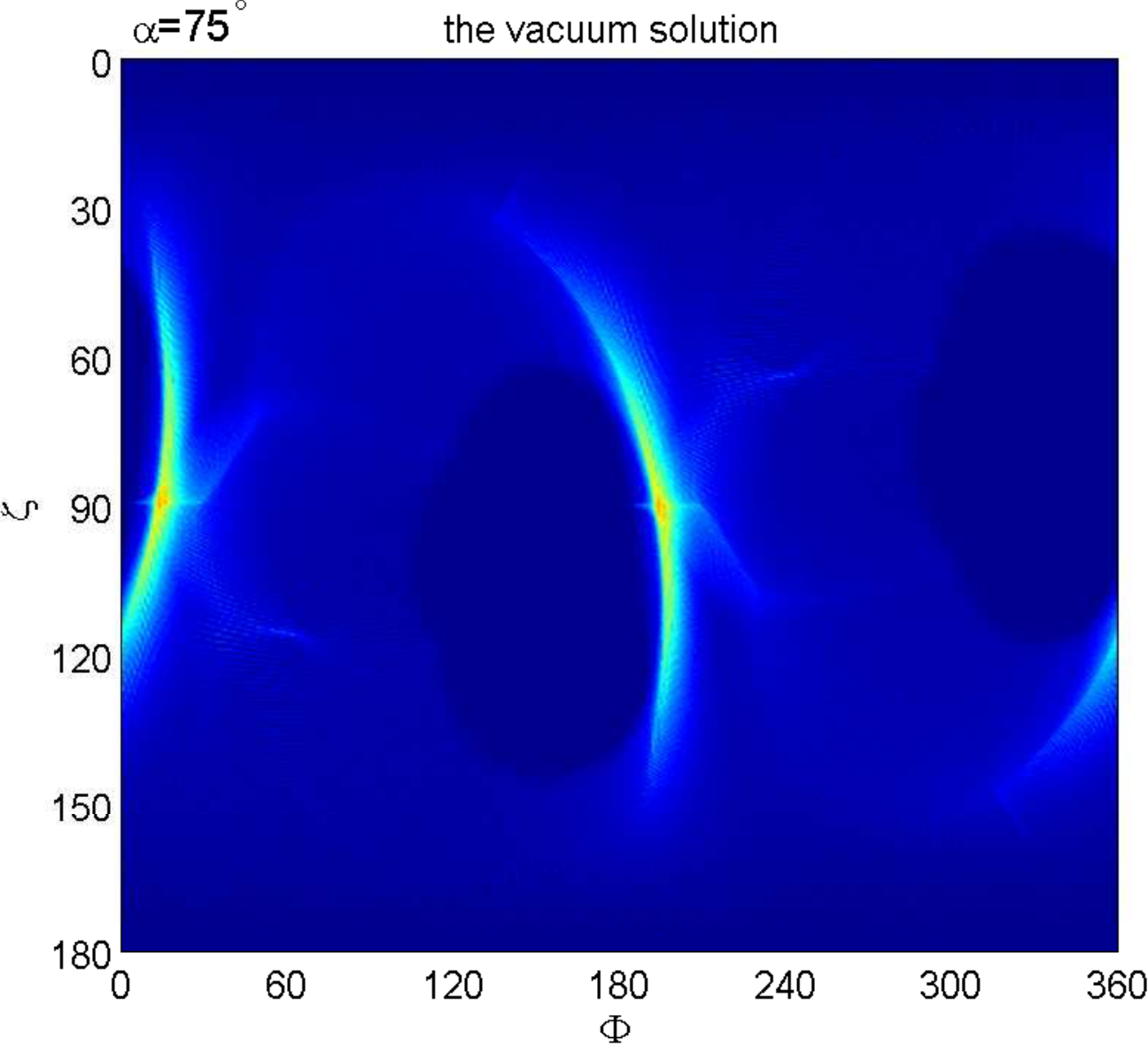} \quad \qquad
  \includegraphics[width=3.5 cm,height=3.75 cm]{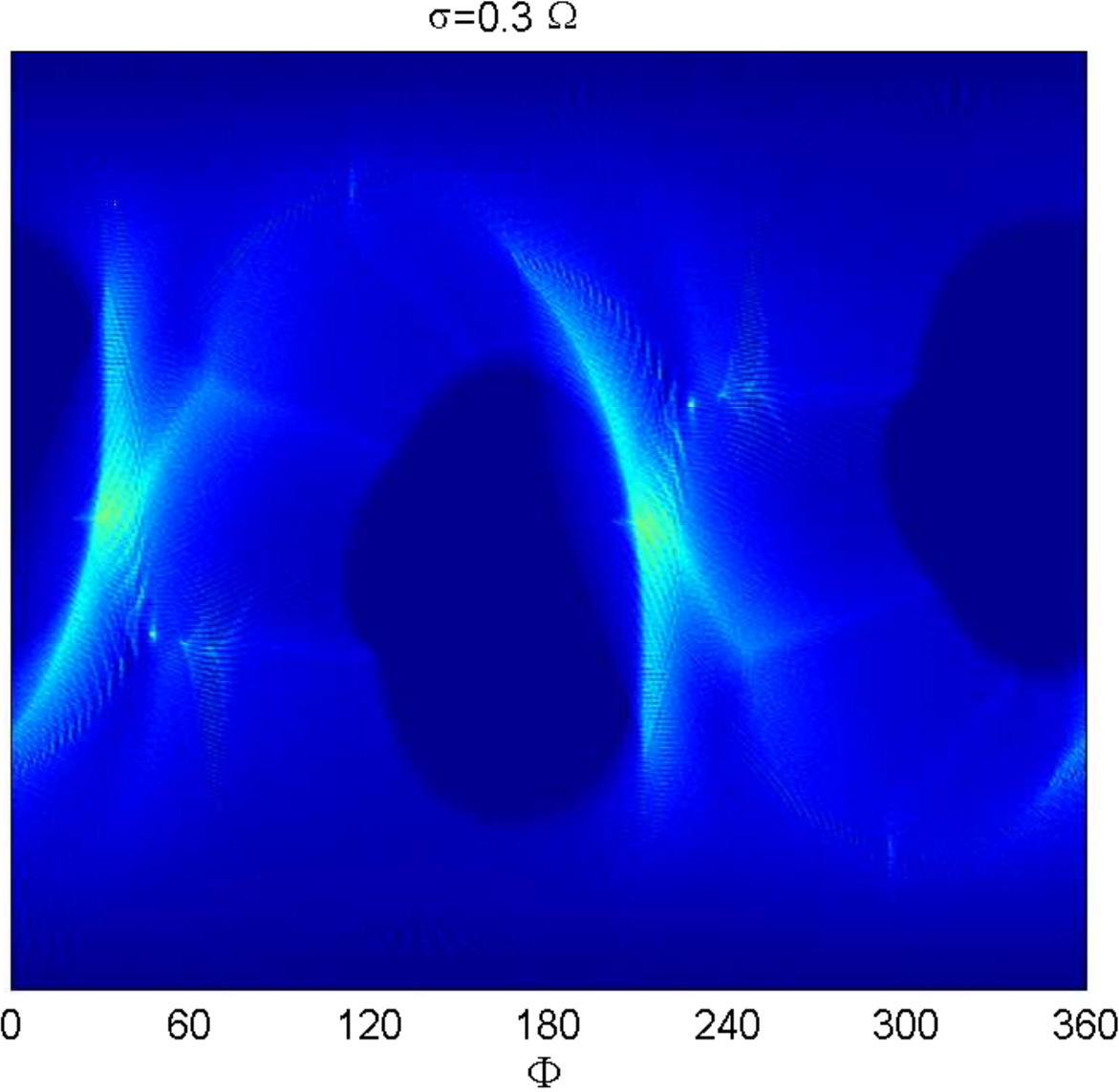} \quad \qquad
  \includegraphics[width=3.5 cm,height=3.75 cm]{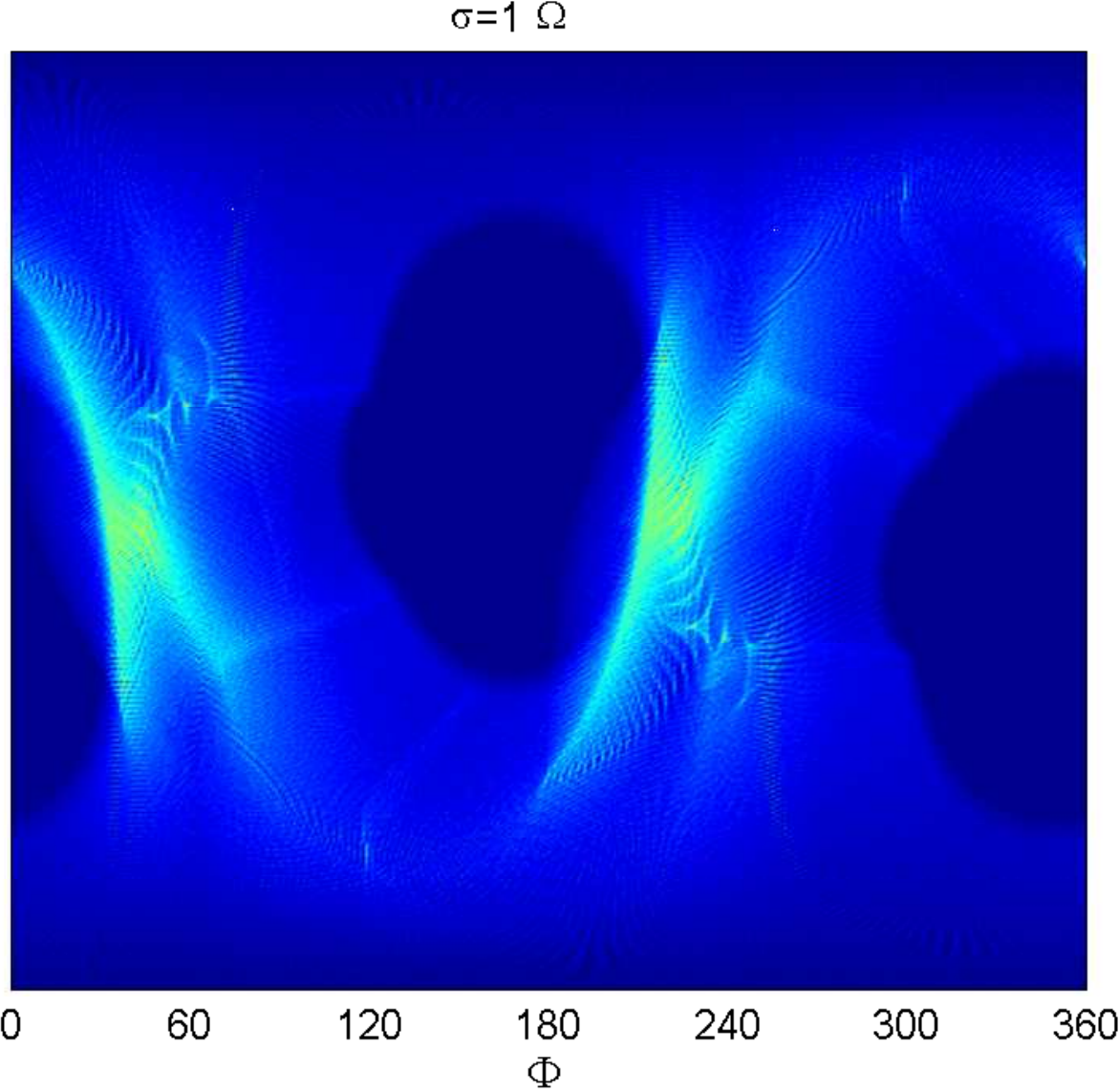} \quad \qquad
  \includegraphics[width=3.5 cm,height=3.75 cm]{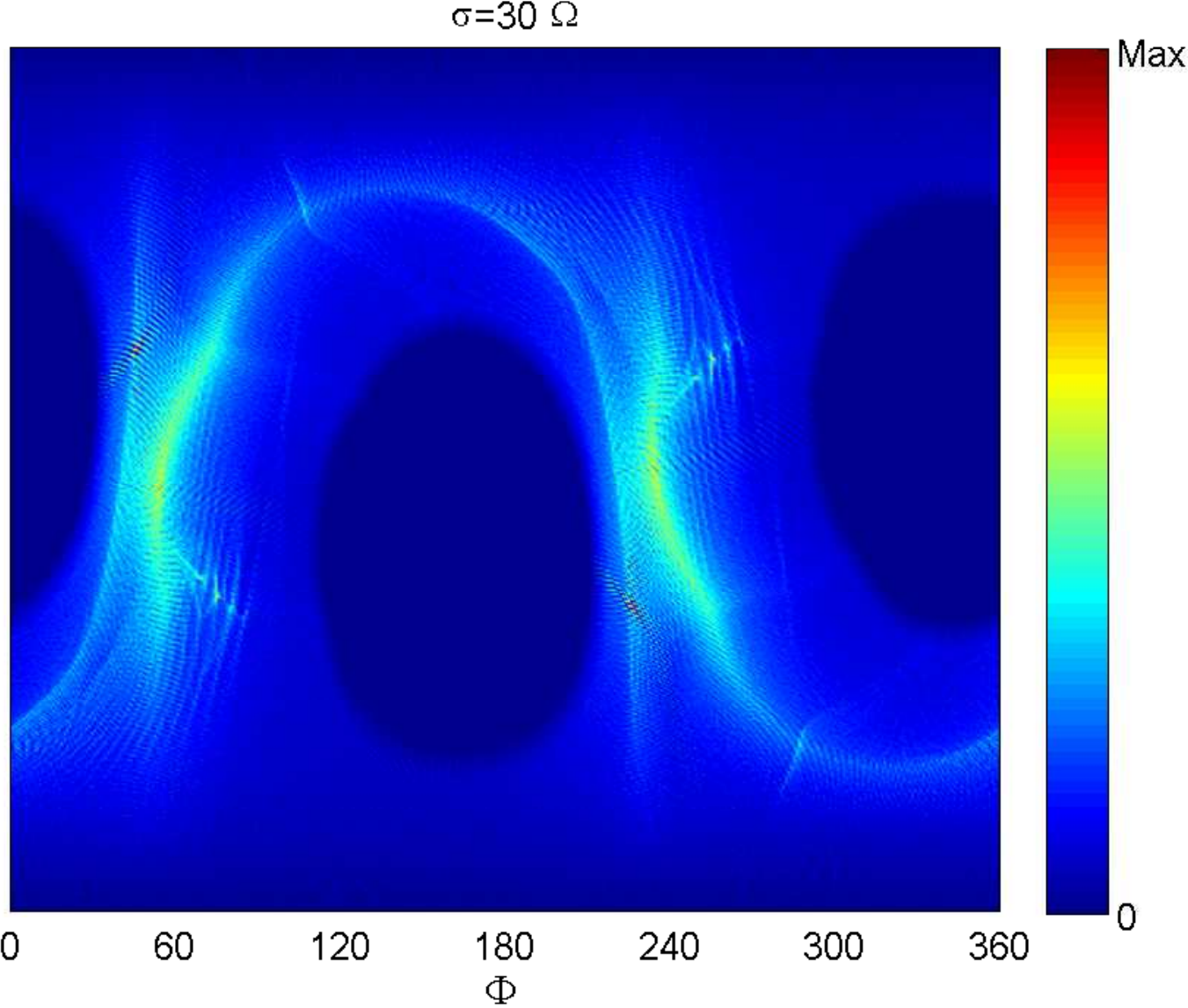} \\\\
  \includegraphics[width=3.75 cm,height=3.8 cm]{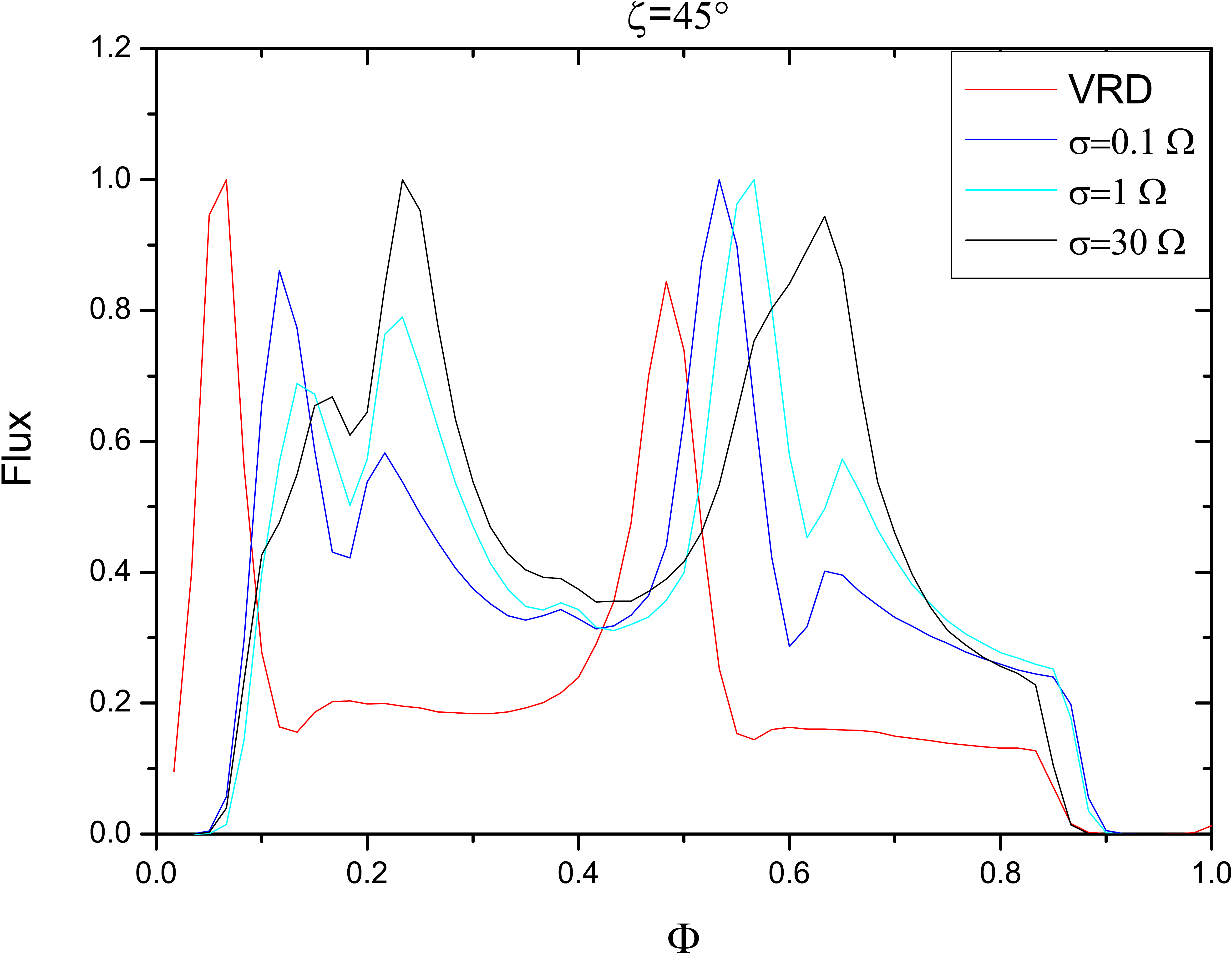} \quad \qquad
  \includegraphics[width=3.5 cm,height=3.8 cm]{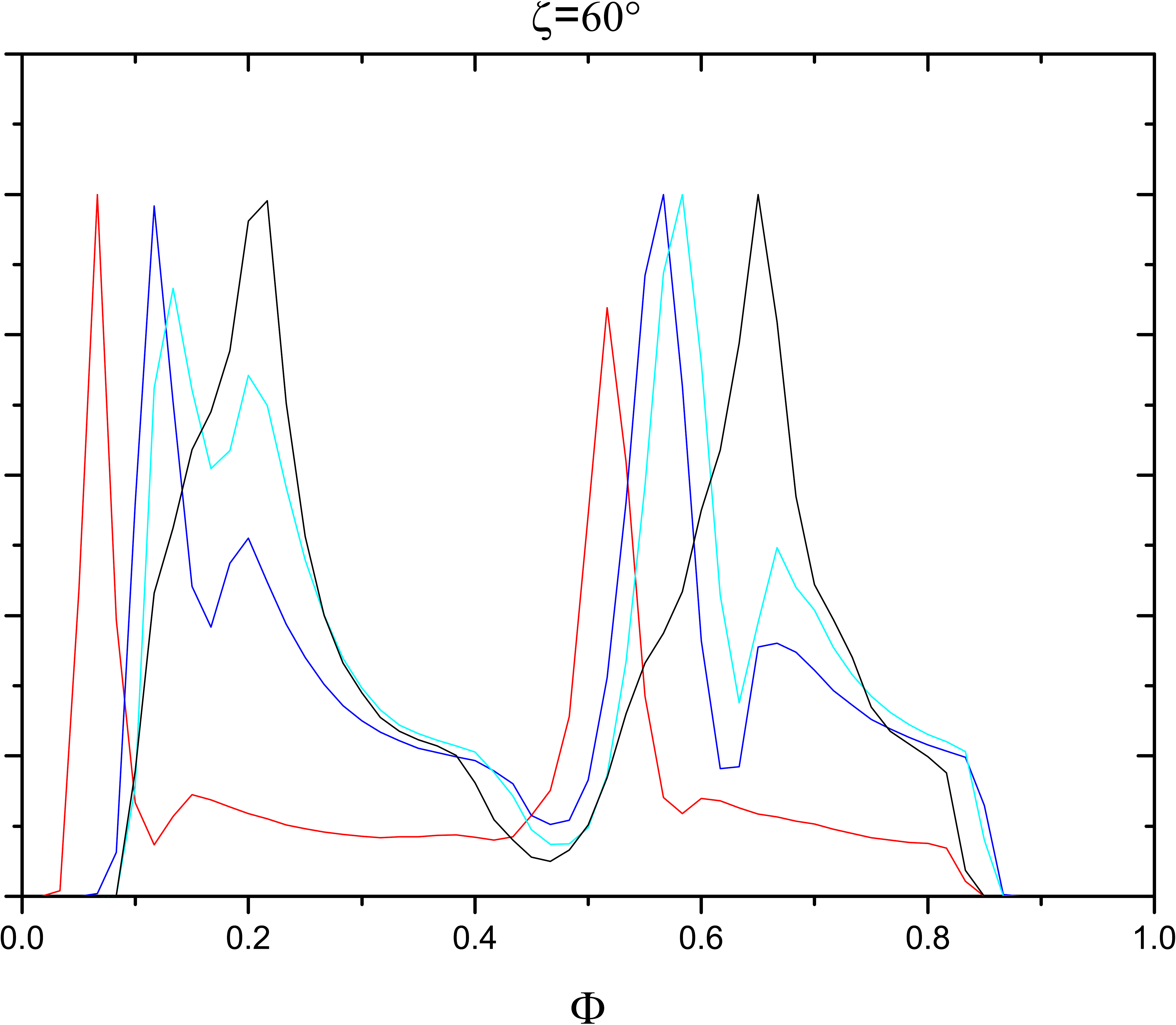} \quad \qquad
  \includegraphics[width=3.5 cm,height=3.8 cm]{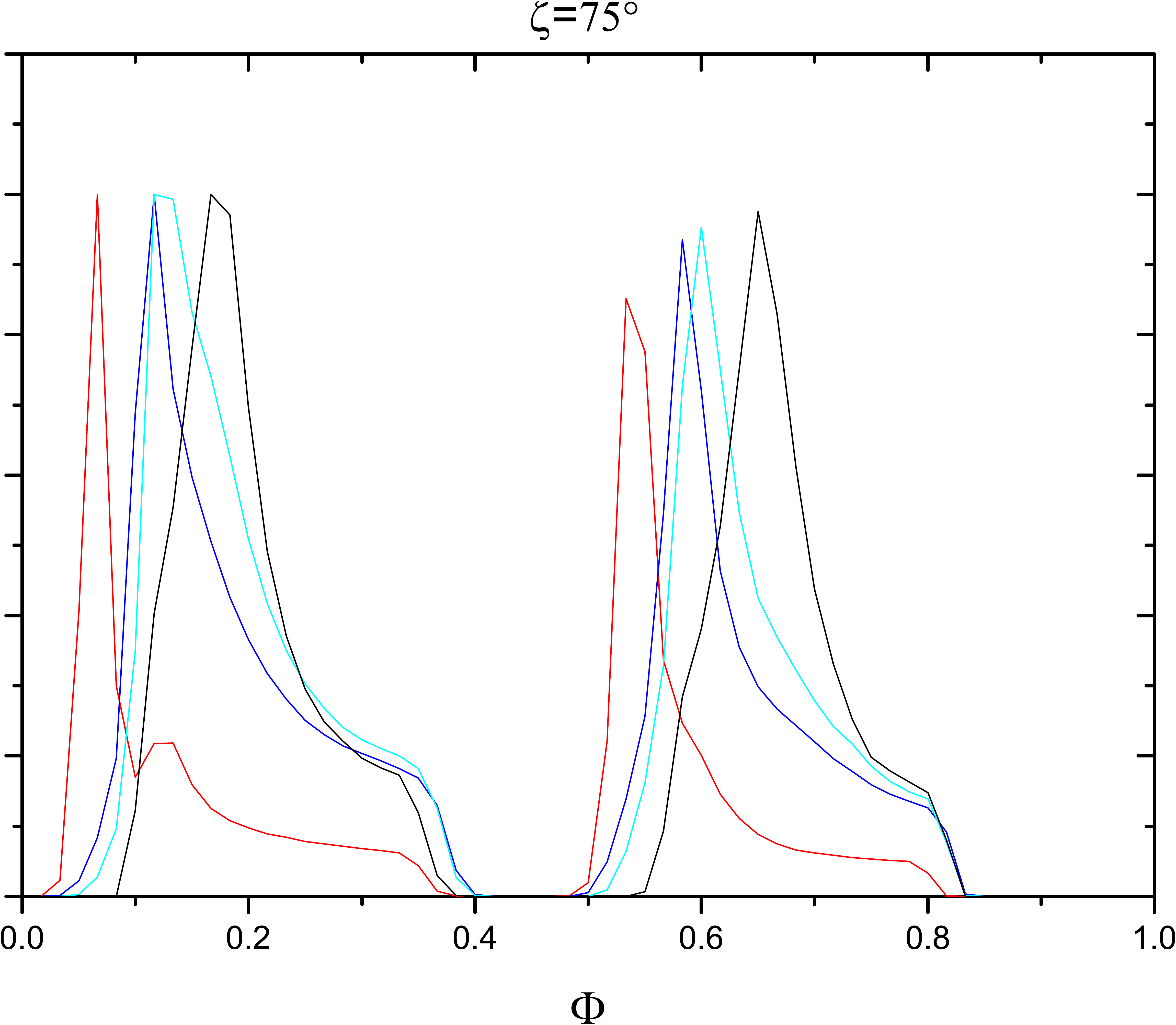} \quad \qquad
  \includegraphics[width=3.5 cm,height=3.8 cm]{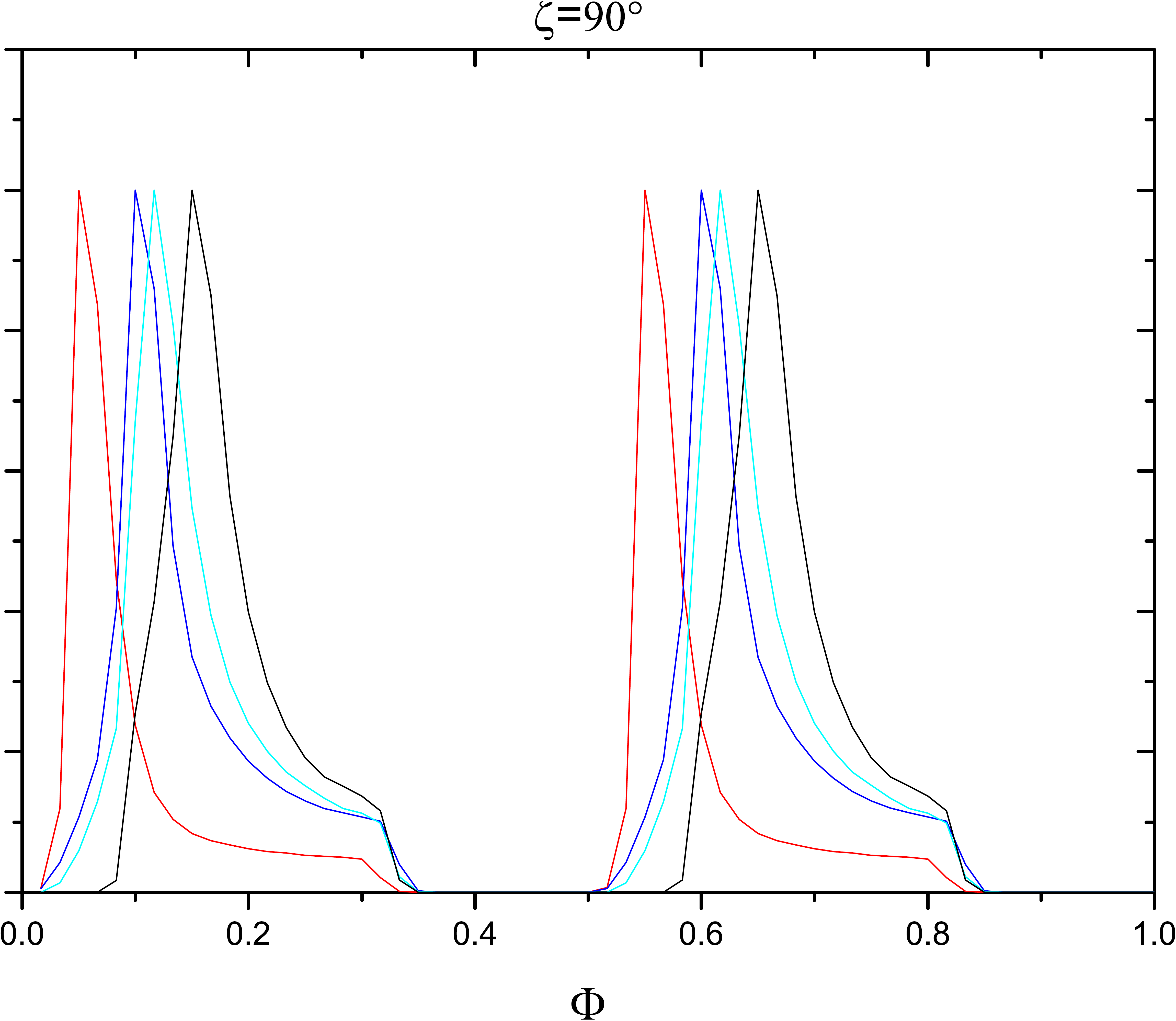} \\\\
  \includegraphics[width=3.75 cm,height=3.75 cm]{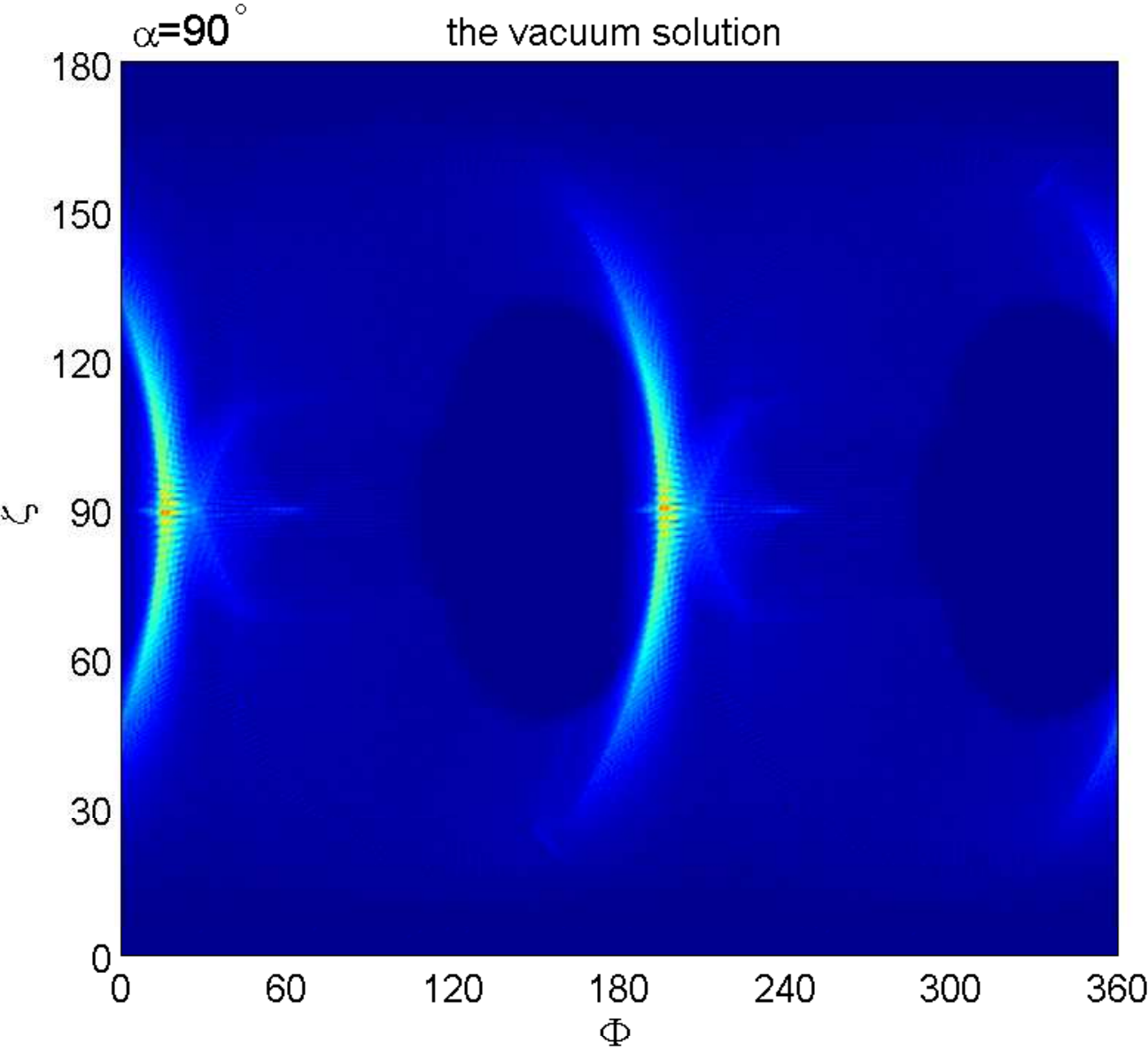} \quad \qquad
  \includegraphics[width=3.5 cm,height=3.75 cm]{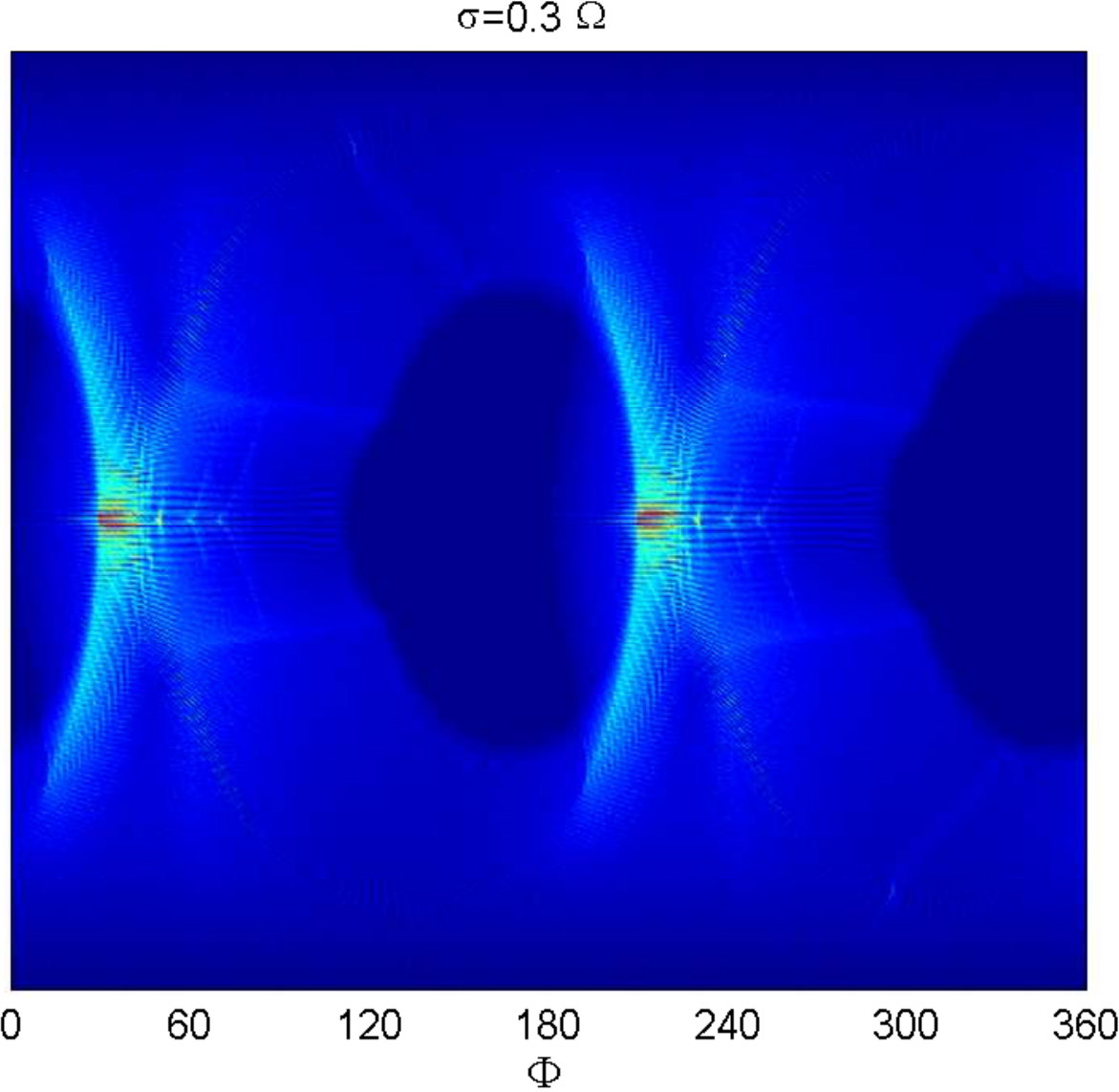} \quad \qquad
  \includegraphics[width=3.5 cm,height=3.75 cm]{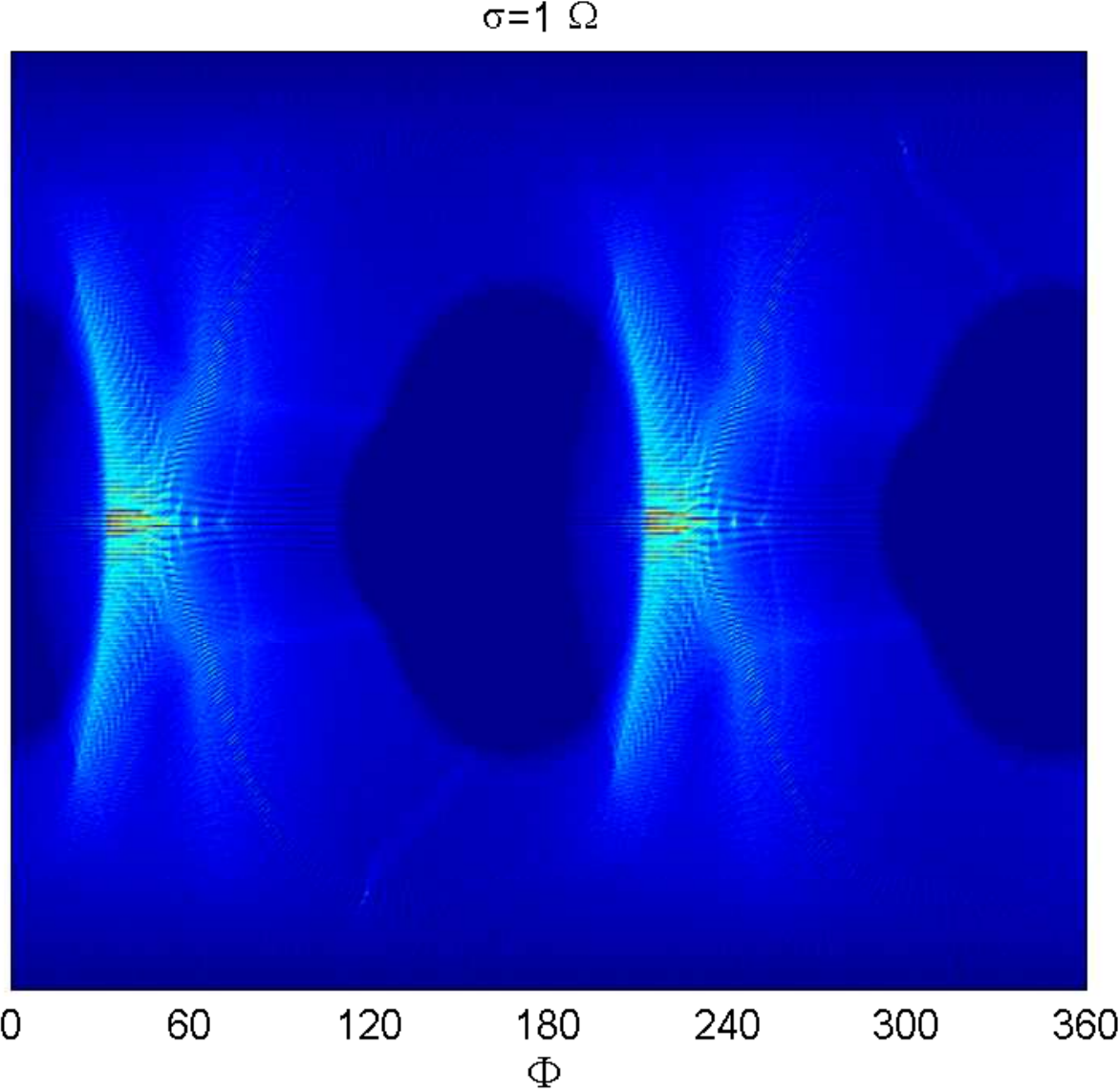} \quad \qquad
  \includegraphics[width=3.5 cm,height=3.75 cm]{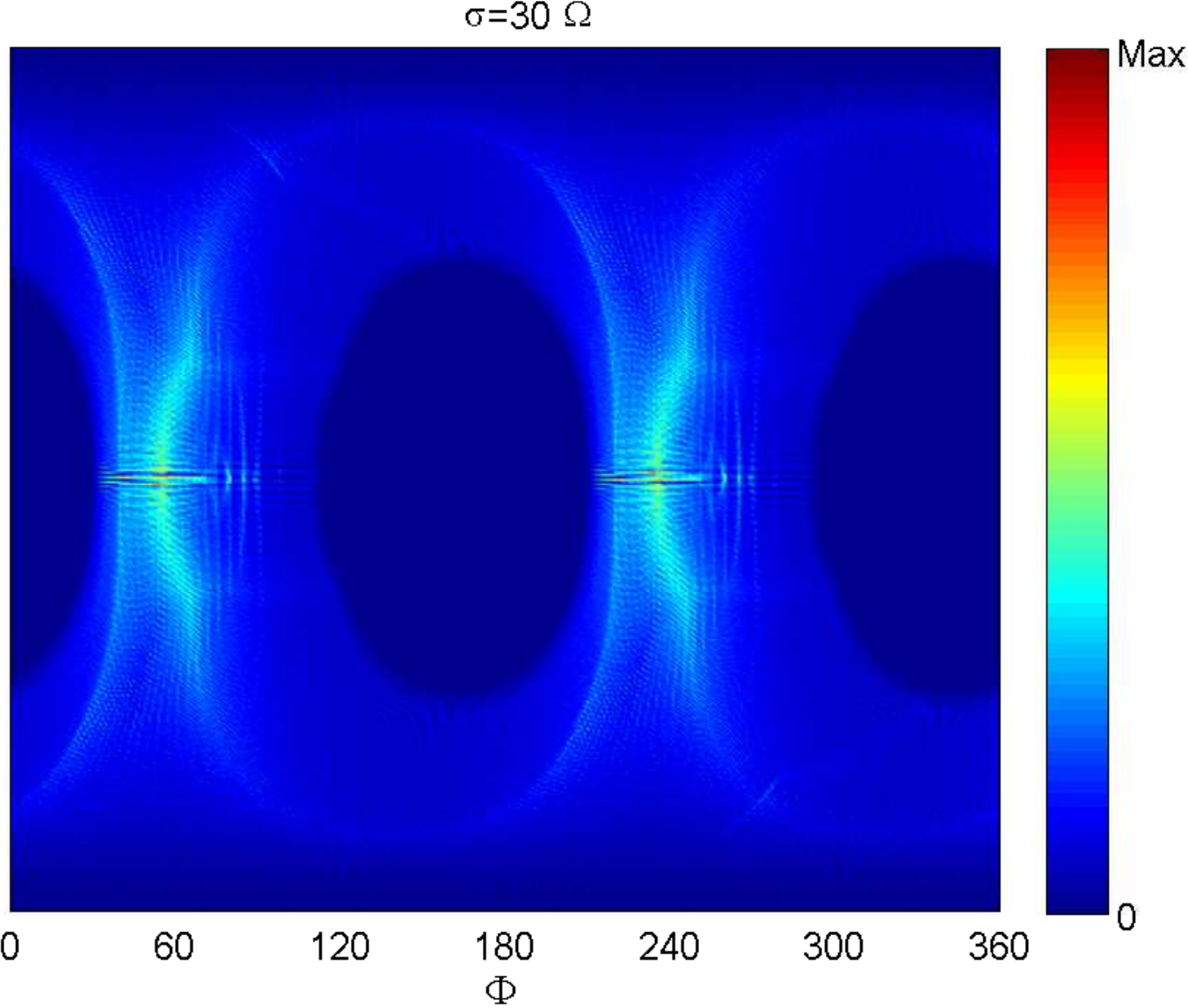} \\\\
  \includegraphics[width=3.75 cm,height=3.8 cm]{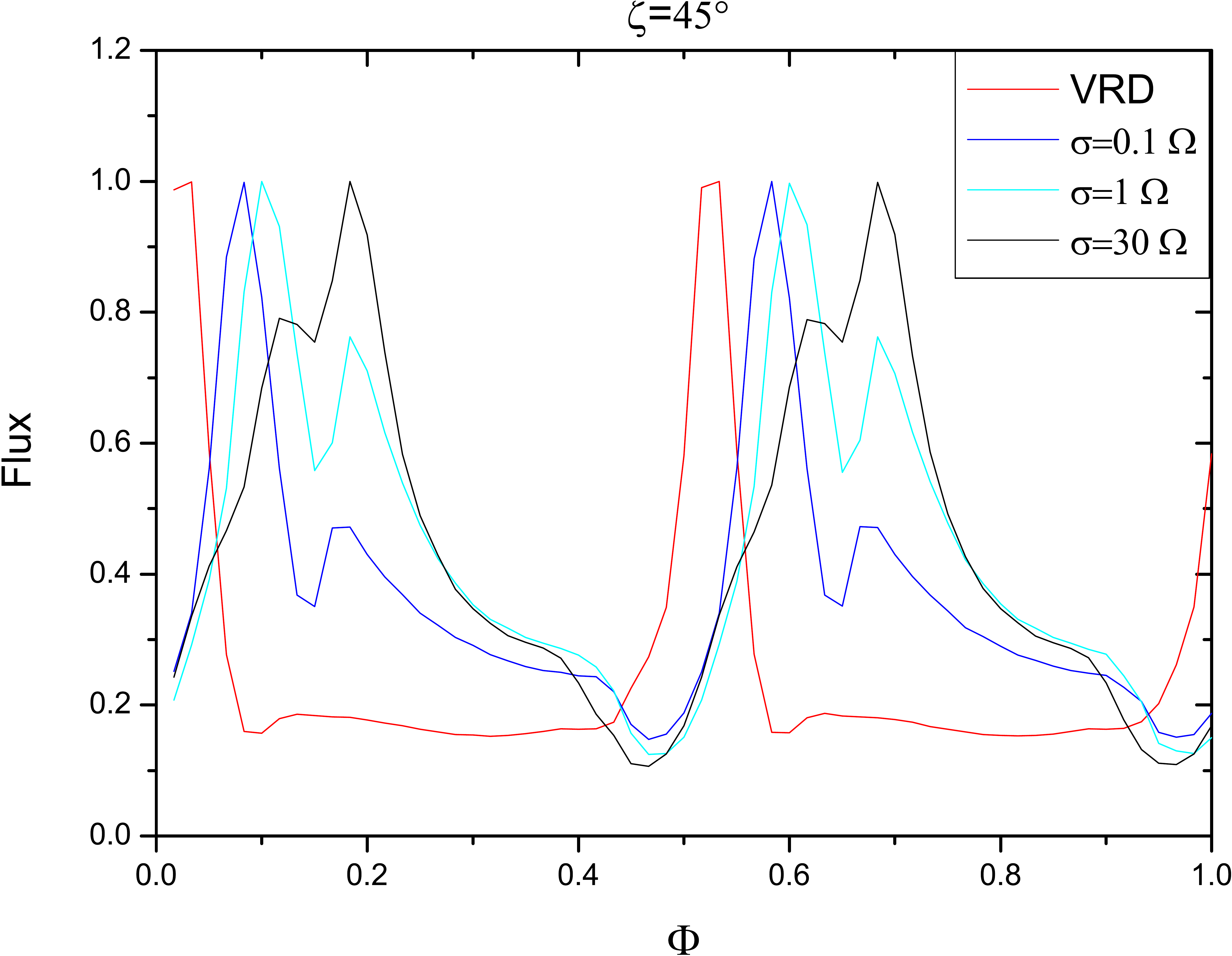} \quad \qquad
  \includegraphics[width=3.5 cm,height=3.8 cm]{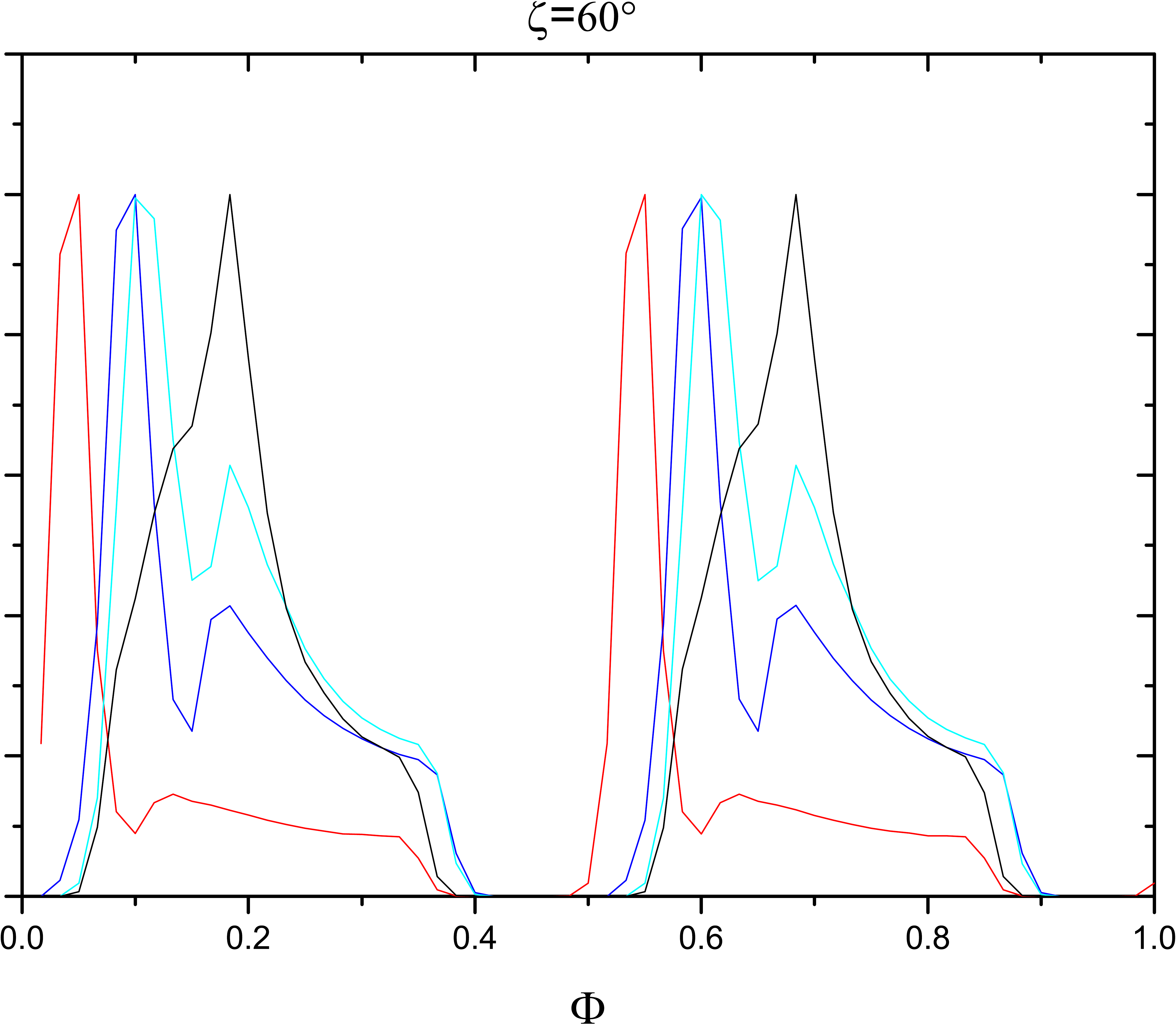} \quad \qquad
  \includegraphics[width=3.5 cm,height=3.8 cm]{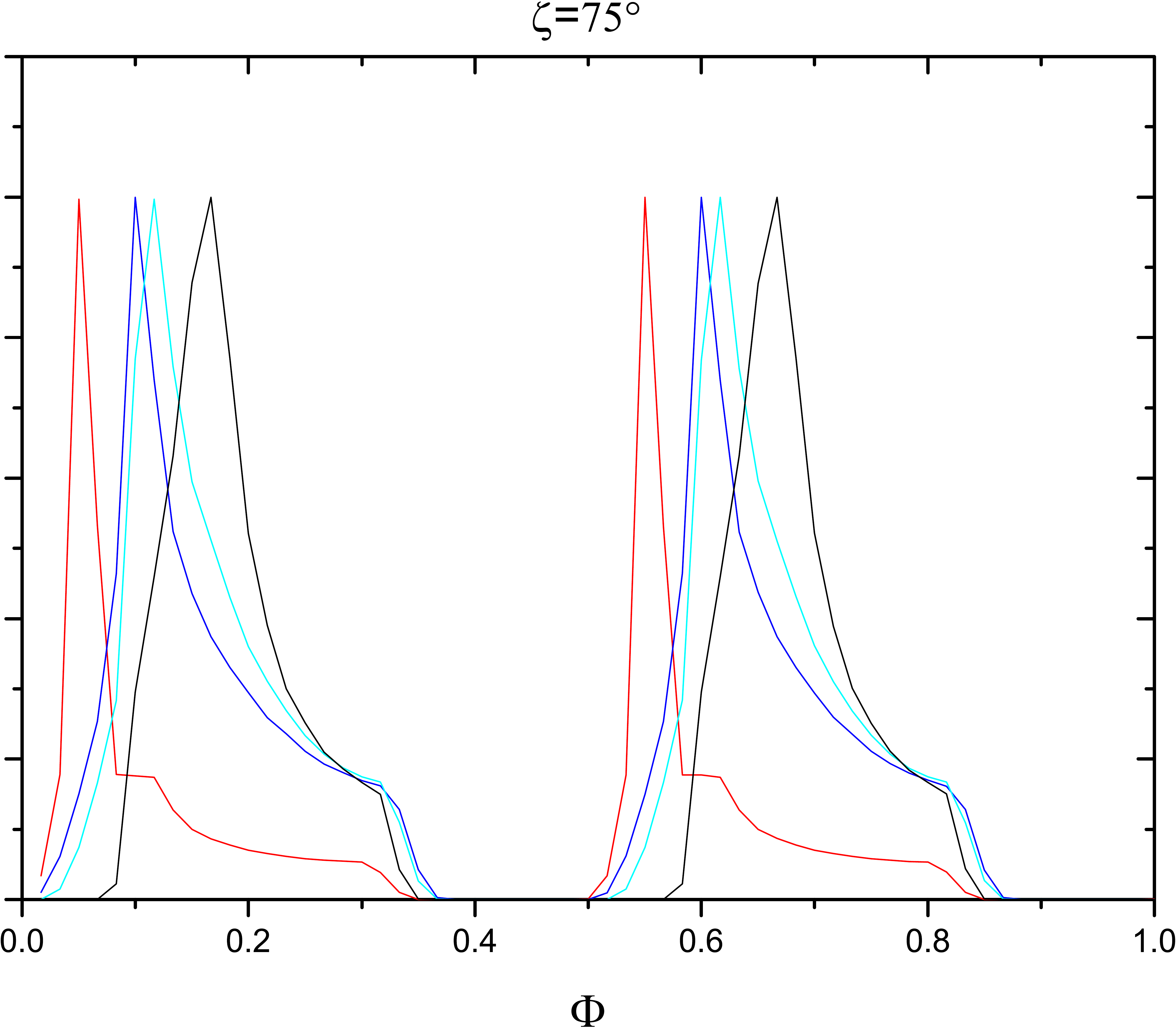} \quad \qquad
  \includegraphics[width=3.5 cm,height=3.8 cm]{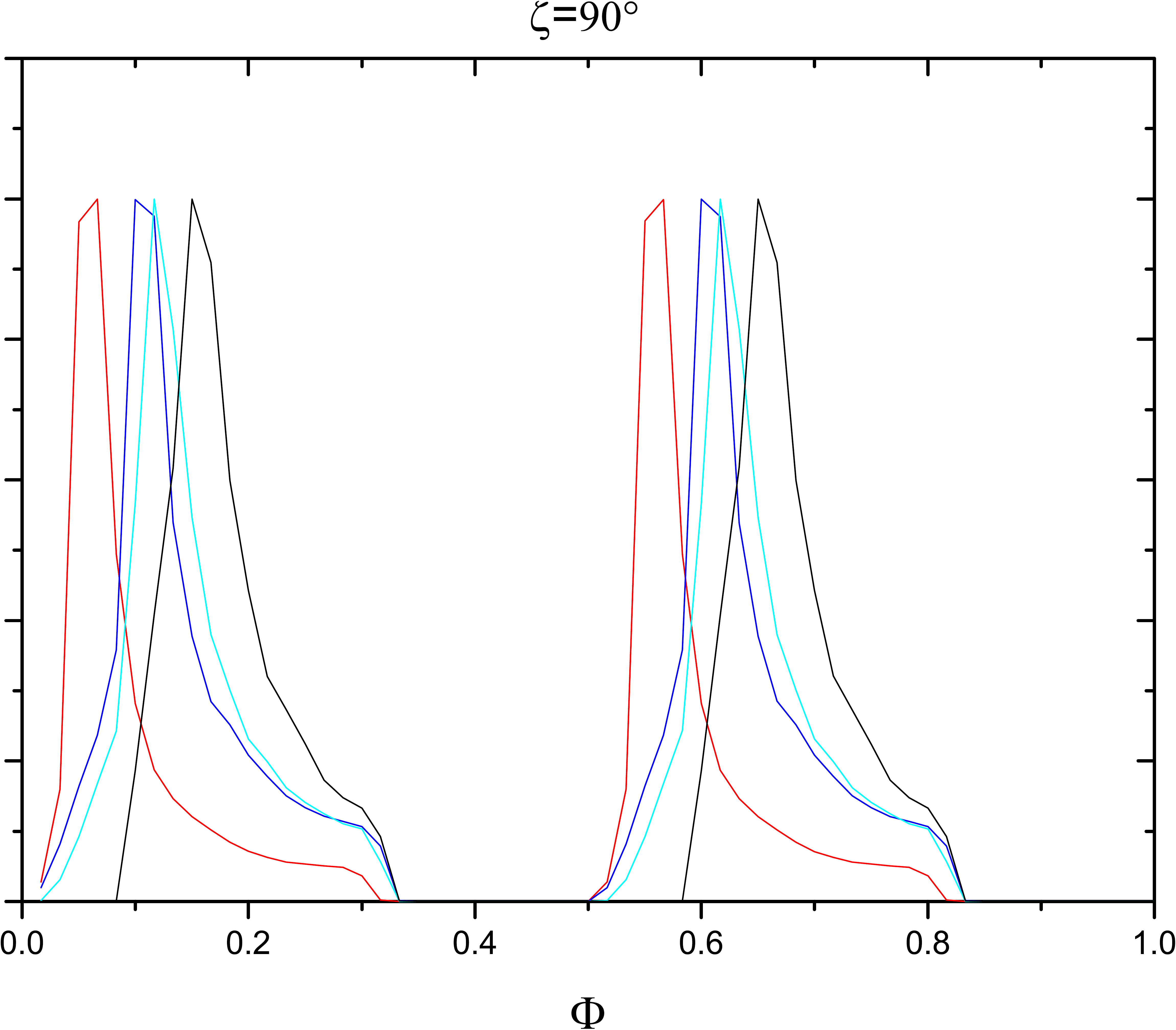}
\end{tabular}
\caption{The sky maps and the corresponding light curves from the geometric method in different inclination angles and view angles with different magnetic field configurations.  }
\end{figure*}

\begin{figure}
\center
\begin{tabular}{cccccc}
\\
\includegraphics[width=7.4cm,height=6.4cm]{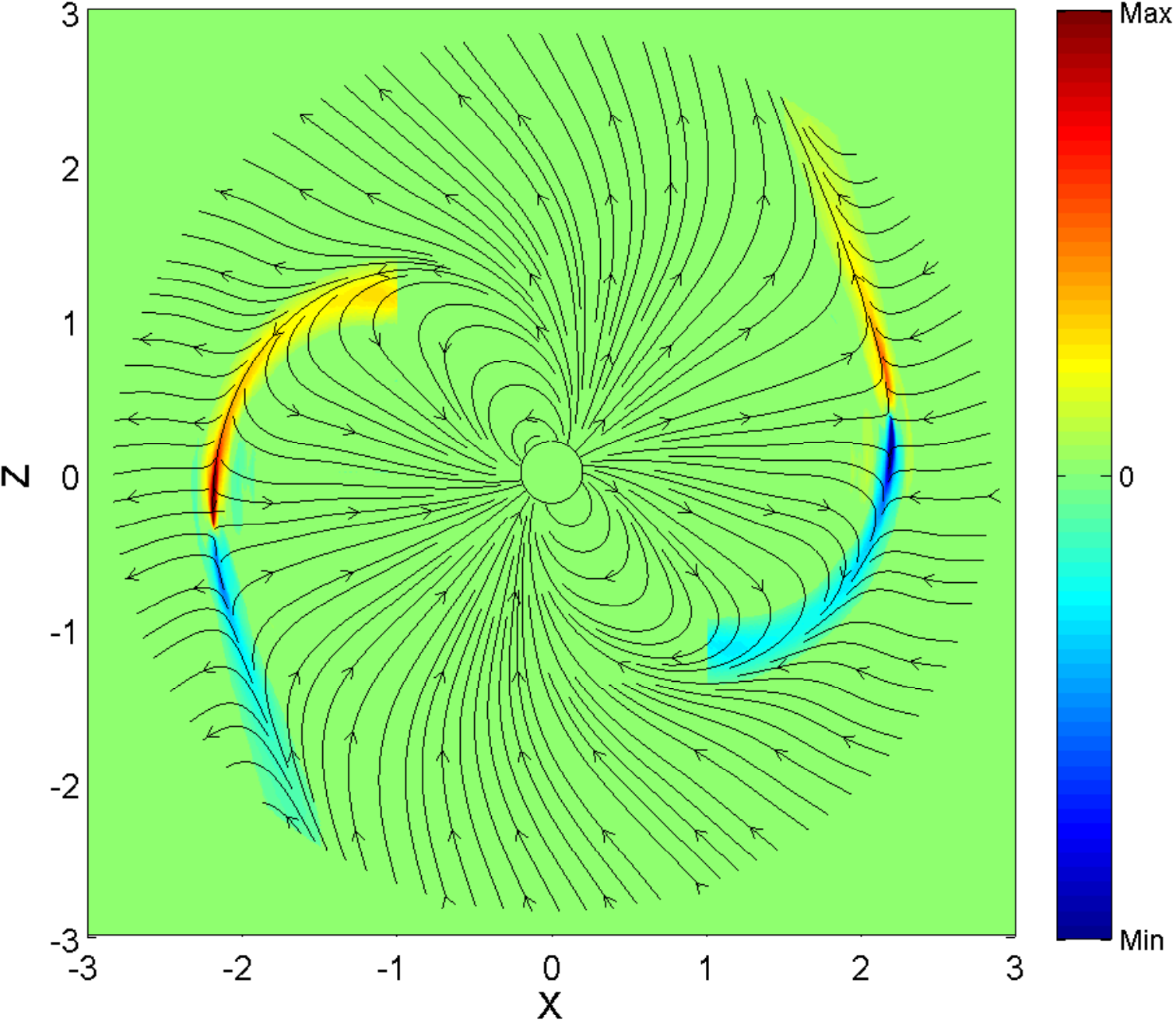}
\end{tabular}
\caption{Distribution of the parallel electric field in the $\Omega-\mu$ plane for magnetic inclination $\alpha=60^{\circ}$ with  $\sigma=60 \, \Omega$ within the LC and with $\sigma=30 \, \Omega$ outside the LC, where any $E_{\|}$ solution is disregarded within the LC. \\ }
\end{figure}

In the previous paper, we presented the structures of oblique pulsar magnetosphere with the uniform conductivity by a pseudo-spectral method \citep{cao16b}. We constructed a set of resistive solutions that smoothly bridges the gap between the vacuum and force-free limits.
In this paper, we focus on the the influence of the conductivity on gamma-ray light curves in dissipation pulsar magnetospheres.
The light curves are produced by the geometric method and the particle trajectory method. As an application, we compare the predicted light curves with the gamma-ray light curves of Vela pulsar observed by Fermi-LAT.
The paper is organized as follows: In section 2, we present the basic equations describing the structure of pulsar magnetosphere. In section 3 and 4, we present the method that we have used to produce the gamma-ray light curves. In section 5, we apply the model to Vela pulsar. Finally, a brief discussion and conclusions are given in section 6.

\section{Basic Equations}
The time-dependent Maxwell equations are given by
\begin{eqnarray}
\frac{1}{c}{\partial {\bf B}\over \partial t}&=&-{\bf \nabla} \times {\bf E}\;,\\
\frac{1}{c}{\partial  {\bf E}\over \partial t}&=&{\bf \nabla} \times {\bf B}-\frac{4\pi}{c}{\bf J}\;,
\end{eqnarray}
with two initial conditions
\begin{eqnarray}
\nabla\cdot{\bf B}&=&0\;,\\
\nabla\cdot{\bf E}&=&4\pi \rho_{\rm e}\;,
\label{Eq3-4}
\end{eqnarray}
where $\rho_{\rm e}$ is the charge density and ${\bf J}$ is the current density.  The structure of pulsar magnetosphere can be determined by the prescription for the current density ${\bf{J}}$.
For the force-free magnetosphere, the current density can be derived by force-free condition ($\bf E\cdot \bf B=0$) and the Maxwell equations as \citep{gru99,bla02}
\begin{eqnarray}
{\bf J}= c\rho_{\rm e} {{\bf E} \times {\bf B} \over B^2}+\frac{c}{4\pi}
{({\bf B}\cdot {\bf \nabla}\times {\bf B}-{\bf E}\cdot {\bf \nabla}\times {\bf E}){\bf{B}}
\over B^2}\;.
\label{Eq6}
\end{eqnarray}
which implies that the electric field parallel to the magnetic field component ${\bf {E}}_{\|}$ will be zero.

The force-free approximations do not allow for any particle acceleration along the magnetic field, ${\bf {E}}_{\|}=0$.
Therefore, they do not account for the particle acceleration and pulsed emission in the magnetosphere. More realistic pulsar magnetosphere should allow for ${\bf {E}}_{\|} \neq 0 $ in some regions of the magnetosphere. The resistive magnetosphere can allow for a non-zero parallel electric field component by involving a conductivity parameter $\sigma$. In fact, the resistive pusar plasma is effectively collisionless. Therefore, the resistivity coming from coulomb collisions can be practically neglected.  We can define the current density for the resistive magnetosphere by a form of Ohm's law given by  \citep{kal14}
\begin{eqnarray}
{\bf J}= c \rho_{\rm e} {{\bf E} \times {\bf B} \over B^2+E^2_{0}}+\sigma {\bf {E}}_{\|}\;,
\label{Eq6}
\end{eqnarray}
where
\begin{eqnarray}
B^2_{0}-E^2_{0}&=&{\bf B}^2-{\bf E}^2,\\
E_{0}B_{0}&=&{\bf E}\cdot {\bf B}, \quad E_{0}\geq0.
\end{eqnarray}
The first term in equation (6) is the drift current component perpendicular to ${\bf {B}}$, while the second term controls ${\bf {E}}_{\|}$ by the conductivity $\sigma$.
The $E_{0}$ term in equation (6)  ensures the drift current to be subluminal.
When the conductivity $\sigma$ increases from $\sigma=0$ to $\sigma \rightarrow \infty$, we expect to obtain a set of solutions that smoothly transition from the vacuum field to the force-free field.
It is noted that there is no unique prescription for the current density in the resistive electrodynamics. We checked the field structures by using the prescription for the current density
given by \citet{li12}. Our results are also qualitatively very similar to those of \citet{li12}.

\section{Geometric Method}
\subsection{Magnetic Field Structures}
The time-dependent Maxwell equations are solved by a spectral method in spherical coordinates. The radial coordinate $r$ is expanded into the Chebyshev function. However,  the angle coordinates $\theta$ and $\phi$ are expanded into the vector spherical harmonic functions. A spectral filter is used in all directions  in order to ensure the stability of the algorithm and increase the convergent rate of the solution. For a detailed discussion about the pseudo-spectral algorithm, see \citet{pet12} and \citet{cao16a,cao16b}.
The computational domain extends from the inner boundary $r_{\rm min}=0.2\, r_{\rm L}$ to the outer boundary $r_{\rm max}=3 \, r_{\rm L}$. A resolution of $N_r \times N_{\theta} \times N_{\phi}=128 \times 32 \times 64$ is necessary to get a good accuracy.
The magnetic field is initialized to be an rotating vacuum dipolar with magnetic inclination angle $\alpha=\{0^\circ,15^\circ,30^\circ,45^\circ,60^\circ,75^\circ,90^\circ\}$. We impose inner boundary condition at the stellar surface with a rotating electric field ${\bf {E}} = -( {\bf \Omega } \times {\bf r} ) \times {\bf B}/c$. A characteristic compatibility method
is used in order to prevent the inward reflection from the outer boundary.
We let the system evolve for several rotational periods to reach a final steady solution.
We perform a series of simulations with uniform conductivities $\sigma=\{0.1,\, 0.3,\, 1,\, 3,\, 5,\, 10,\, 30,\, 60 \} \,\, \Omega$. Magnetic field structures in the $\Omega-\mu$ plane for magnetic inclination $\alpha=60^{\circ}$ with increasing conductivity are show in figure 1.
The normalized Poynting flux $L/L_{\rm aligned}$ as a function of radius $r$ for magnetic inclination $\alpha=60^{\circ}$ with different $\sigma$ values at time $t= 3 \, P$ are shown in figure 2. We can see that the Poynting flux increases with increasing conductivity. To demonstrate the convergence of the solution, the normalized Poynting flux for $\sigma=30 \, \Omega$ at time $t= 4 \, P$  is also shown as the red dashed line. The Poynting flux is nearly constant when the time $t> 3 \, P$, the system will relax to the steady-state solution.
In figure 3, we show the normalized Poynting flux $L/L_{\rm aligned}$ as a function of radius $r$ for magnetic inclination $\alpha=0^{\circ}$ with different $\sigma$ values at time $t= 3 \, P$. It is found that our results still have some numerical dissipations due to the low resolution used in the presented 3D simulations. A higher resolution is necessary to resolve more fine plasmoids in the current sheet.  Our simulations only present the lower limit on resolution to get convergence.
We can see a smooth transition for magnetic field structures and the Poyting flux from the vacuum limit to the force-free limit with increasing conductivity.
For a detailed description about the structures of resistive magnetosphere, see \citet{cao16b}.

The polar cap shape is defined by the footprints of the last closed field lines on the stellar surface. The last closed field lines can be found by checking whether the field lines close inside or outside the LC. It has been known that the polar cap shape in vacuum case is different from that in the force-free case \citep[e.g., ][]{bai10b,kal14}. Here, a third-order Runge-Kutta integration is used to find the last closed field lines in the lab frame, and the bisection method is used to find the magnetic colatitude $\theta^{\rm rim}_{m}$ of the rim of the polar cap for the fixed magnetic azimuth $\phi^{\rm rim}_{m}$. Open volume coordinates are be defined by $(r_{\rm ovc},\phi_{\rm m})$ on the polar cap. $r_{\rm ovc}\geq1$ represent the closed field lines and $r_{\rm ovc}<1$ represent the open field lines. The polar cap shapes on a sphere of radius 0.2\,$r_{\rm L}$ for magnetic inclination $\alpha=60^{\circ}$ with different values of $\sigma$ are shown in figure 4. The polar cap shapes for the low $\sigma$ value are similar to that for the vacuum field, and a notch appears on the polar cap rim. In the resistive solution, on the one hand, the polar shapes are larger than that of vacuum solution because of the larger open magnetic flux. On the other hand, the polar caps are shifted towards the trailing side. The notches on the polar caps disappear at high $\sigma$ value and the polar cap (PC) shapes are more circular.
\begin{figure*}
\begin{tabular}{cccccccc}
\includegraphics[width=4. cm,height=3.5 cm]{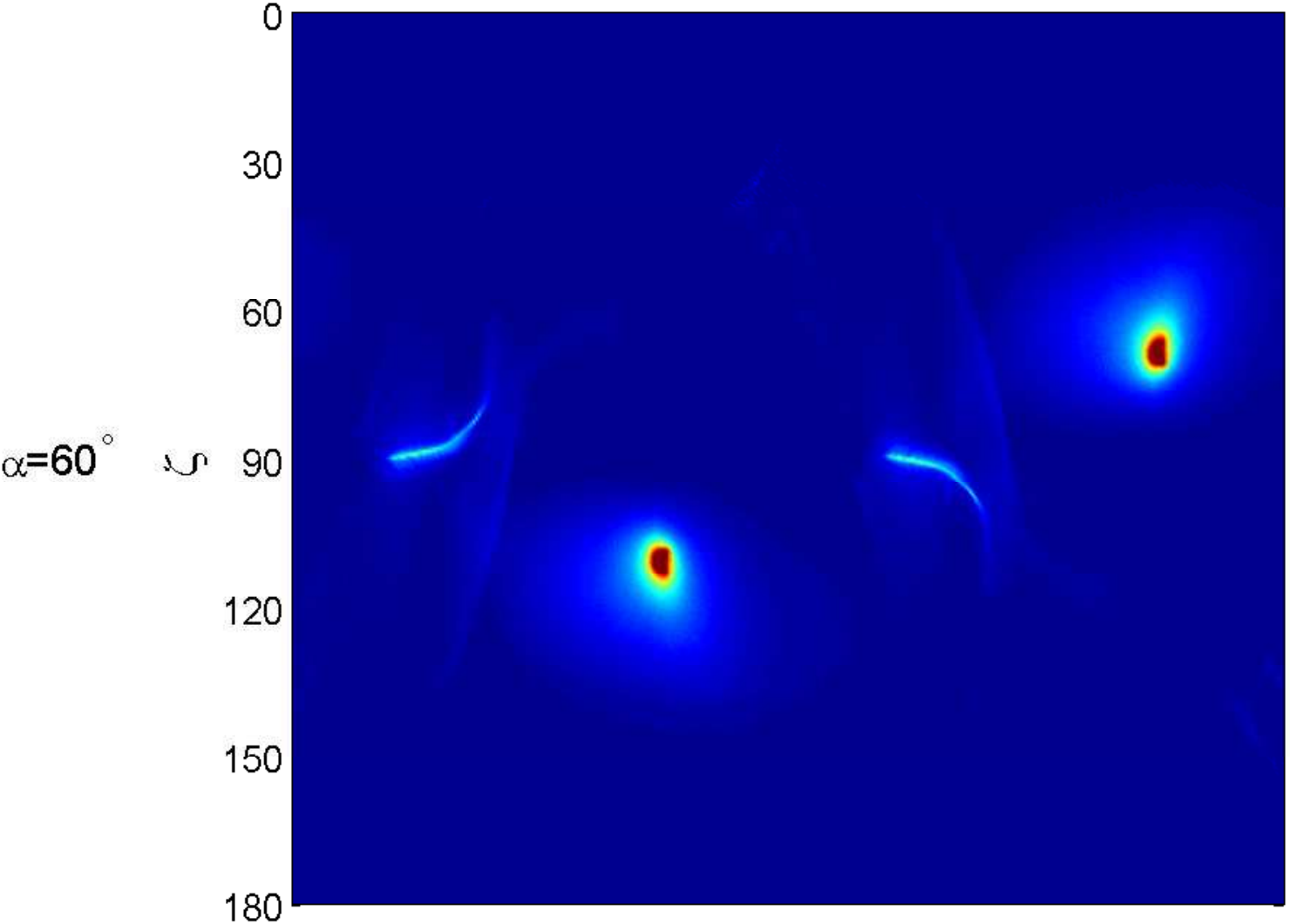} \,
\includegraphics[width=14. cm,height=4. cm]{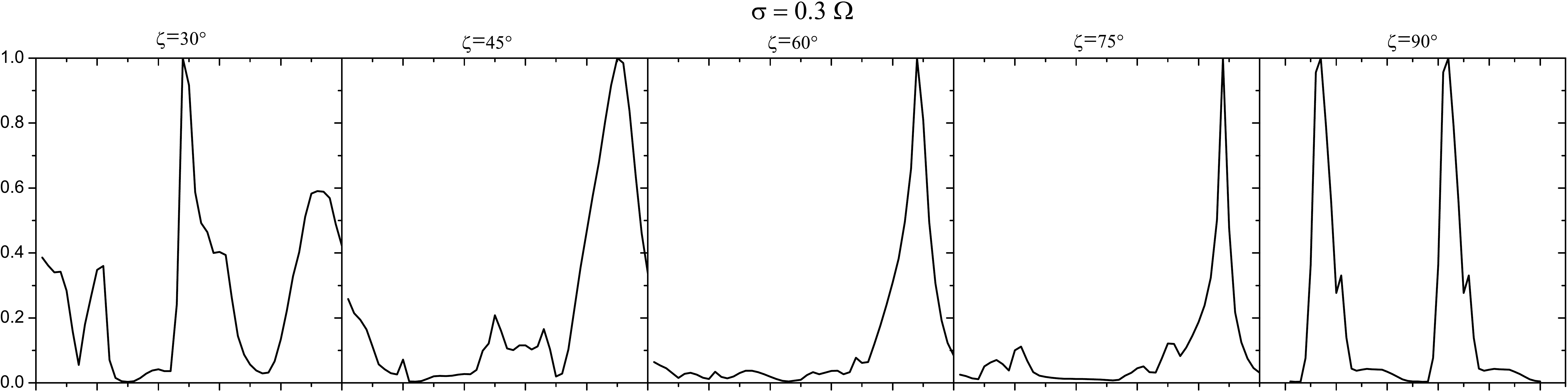} \\ \\
\includegraphics[width=4. cm,height=3.5 cm]{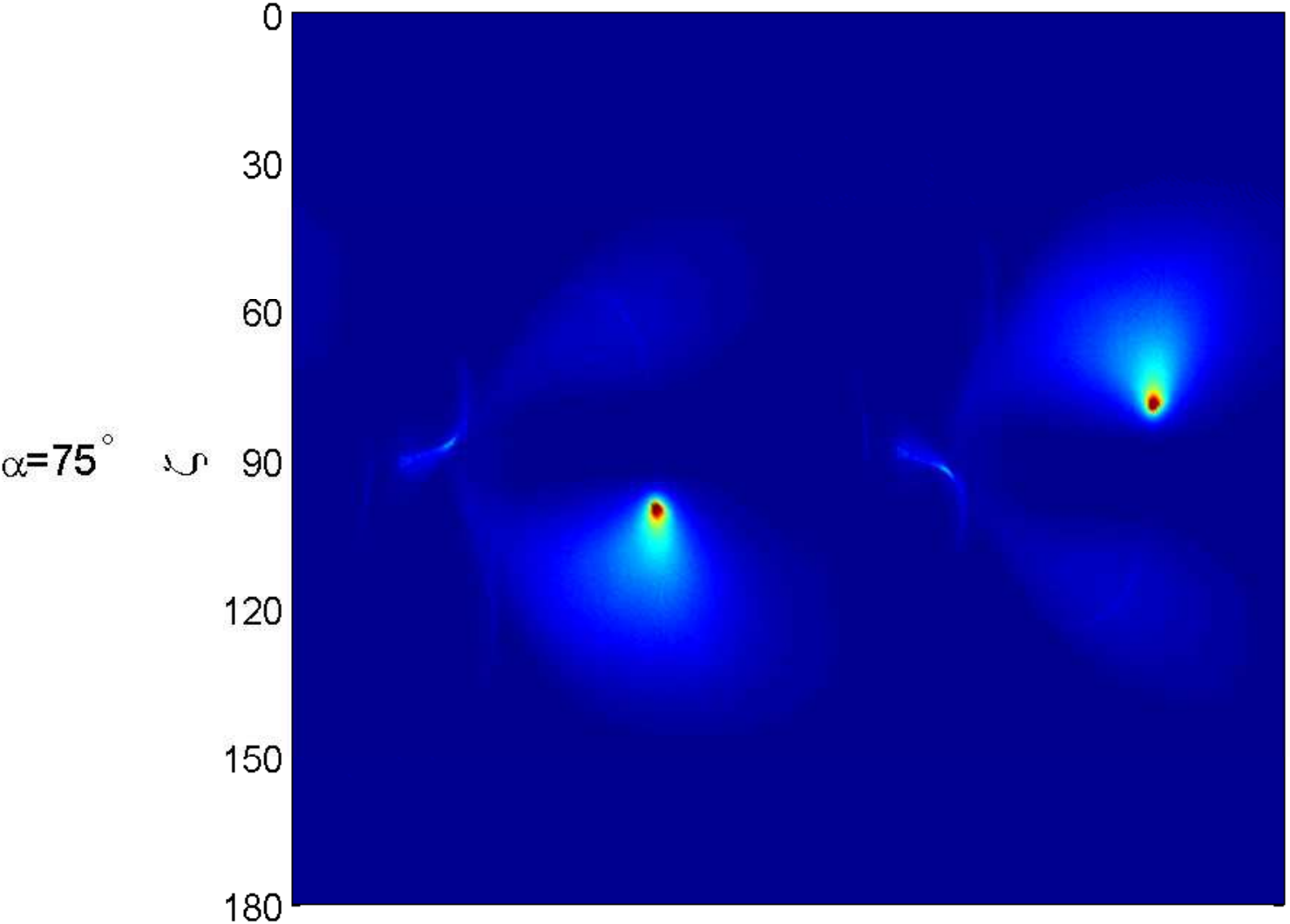} \,
\includegraphics[width=14. cm,height=3.5 cm]{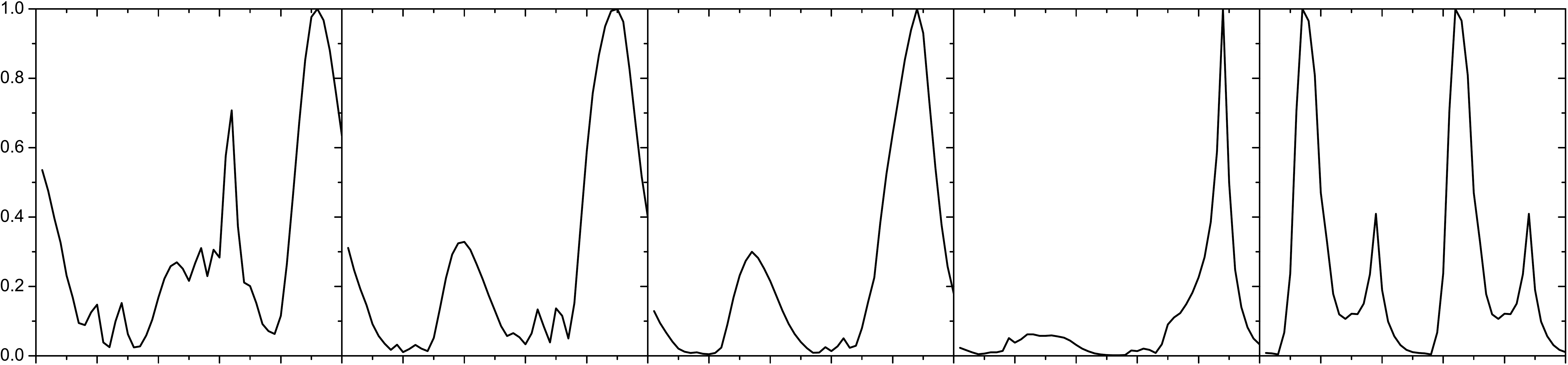} \\ \\
\includegraphics[width=4.1 cm,height=4. cm]{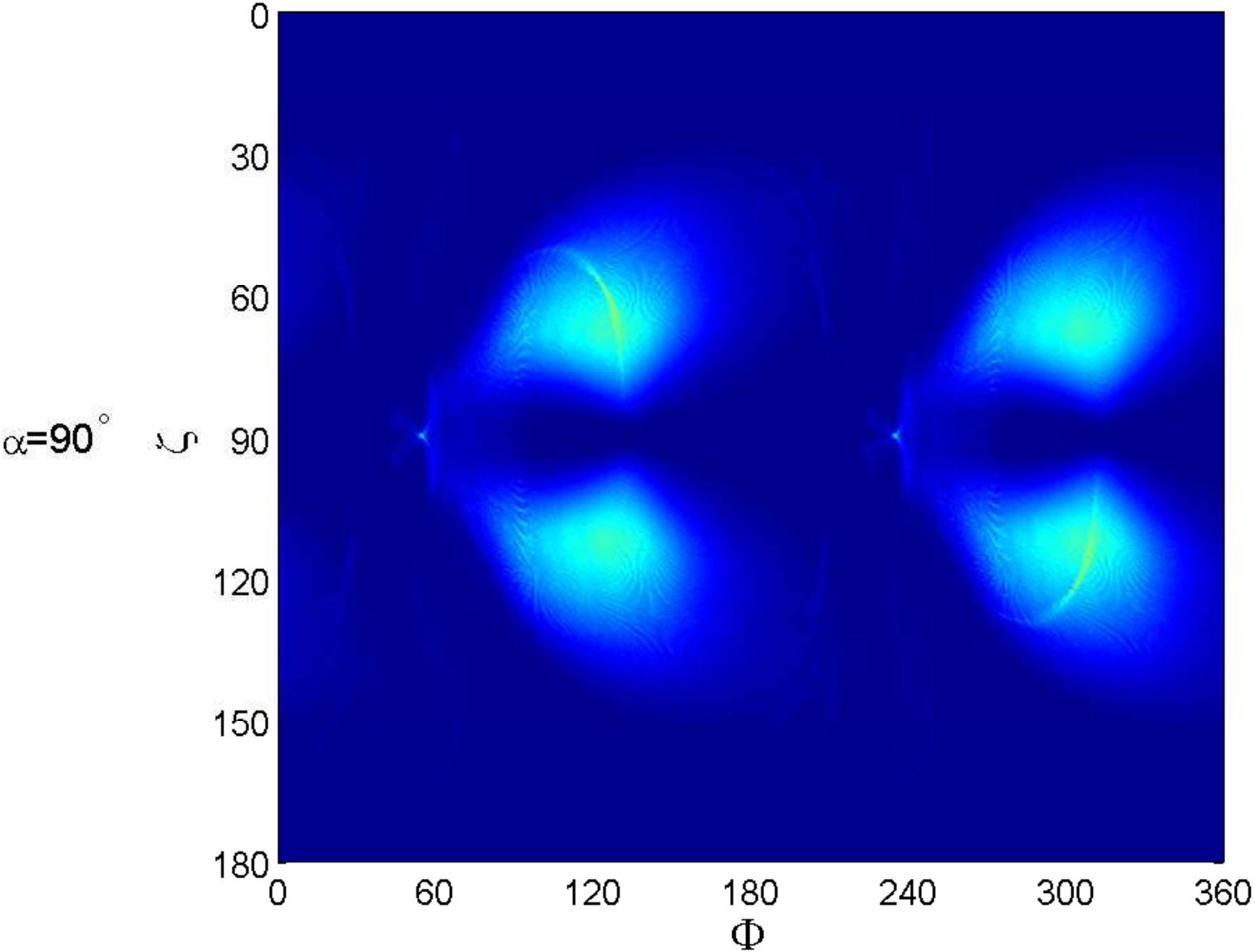} \,
\includegraphics[width=14. cm,height=4. cm]{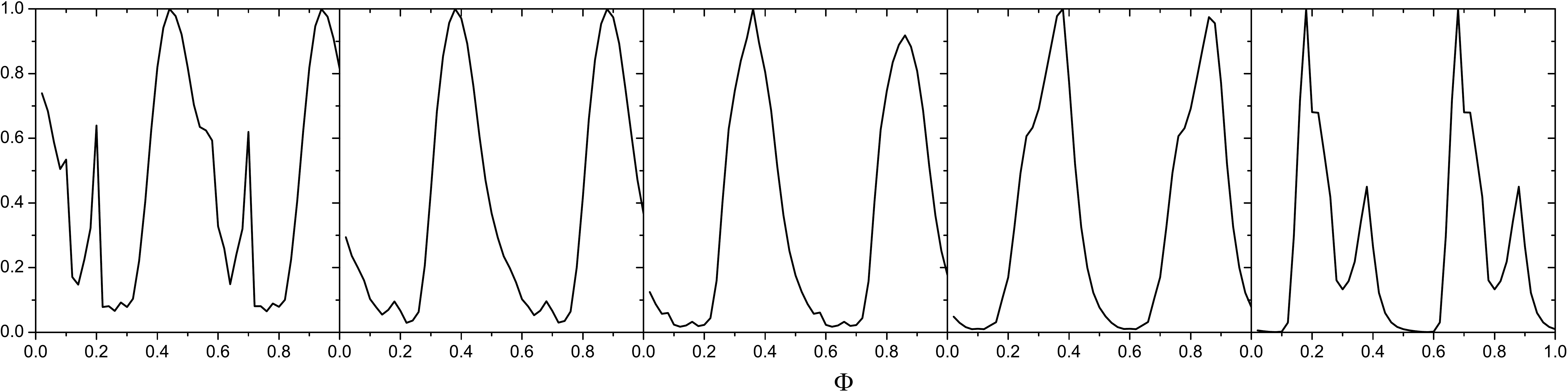} \\ \\
\end{tabular}
\caption{The sky maps and the corresponding light curves from the particle trajectory method in different inclination angles and view angles
for the magnetosphere with a near force-free regime within the LC and with $\sigma=0.3 \, \Omega$ outside the LC.  }
\end{figure*}

\begin{figure*}
\begin{tabular}{cccccccc}
\includegraphics[width=4. cm,height=3.5 cm]{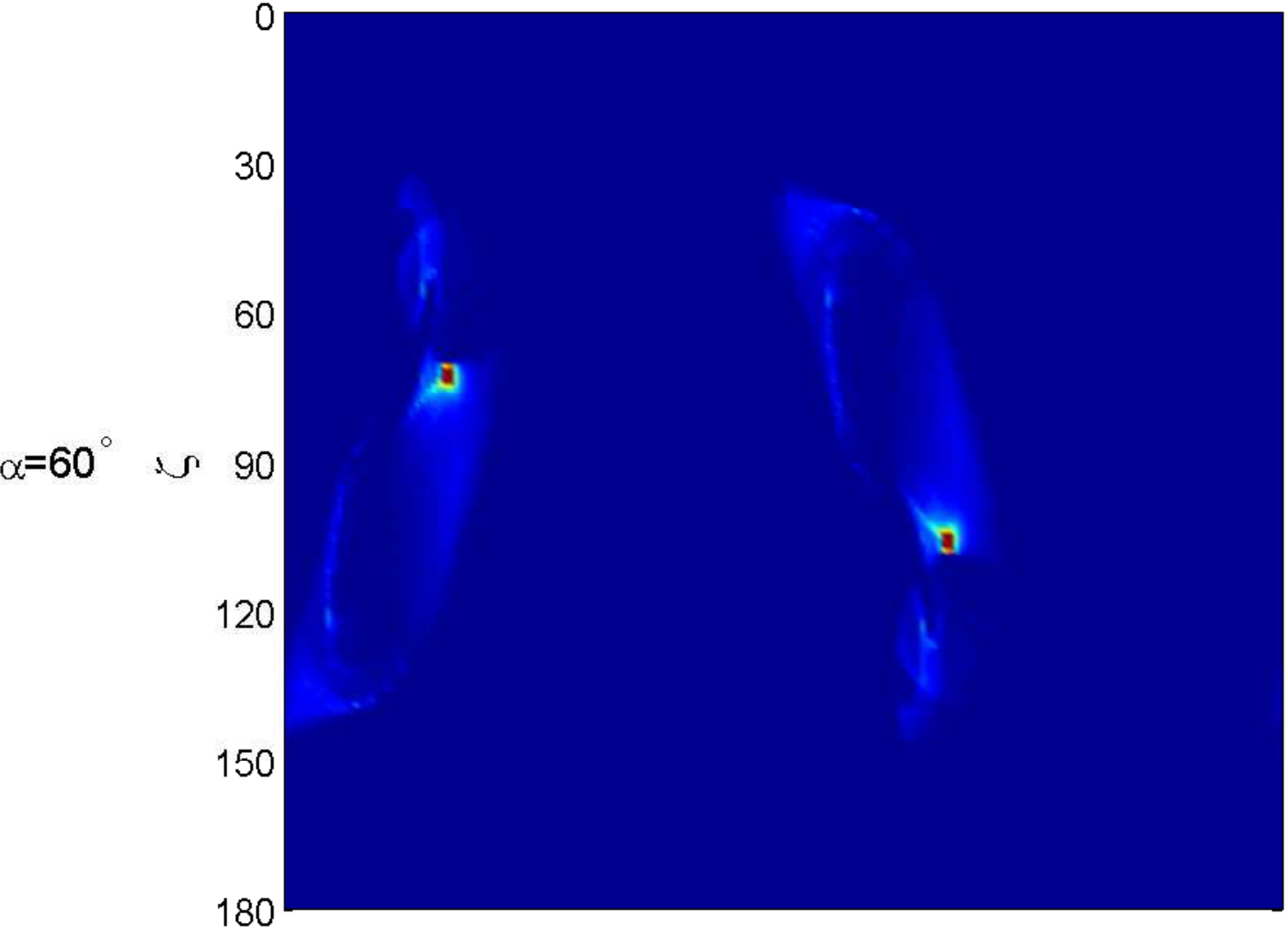} \,
\includegraphics[width=14. cm,height=4. cm]{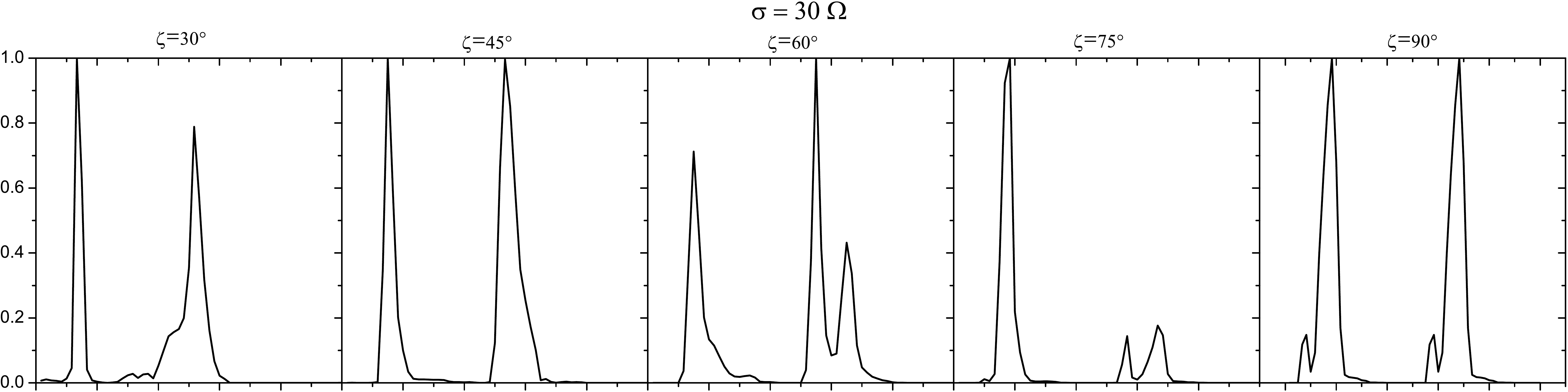} \\ \\
\includegraphics[width=4. cm,height=3.5 cm]{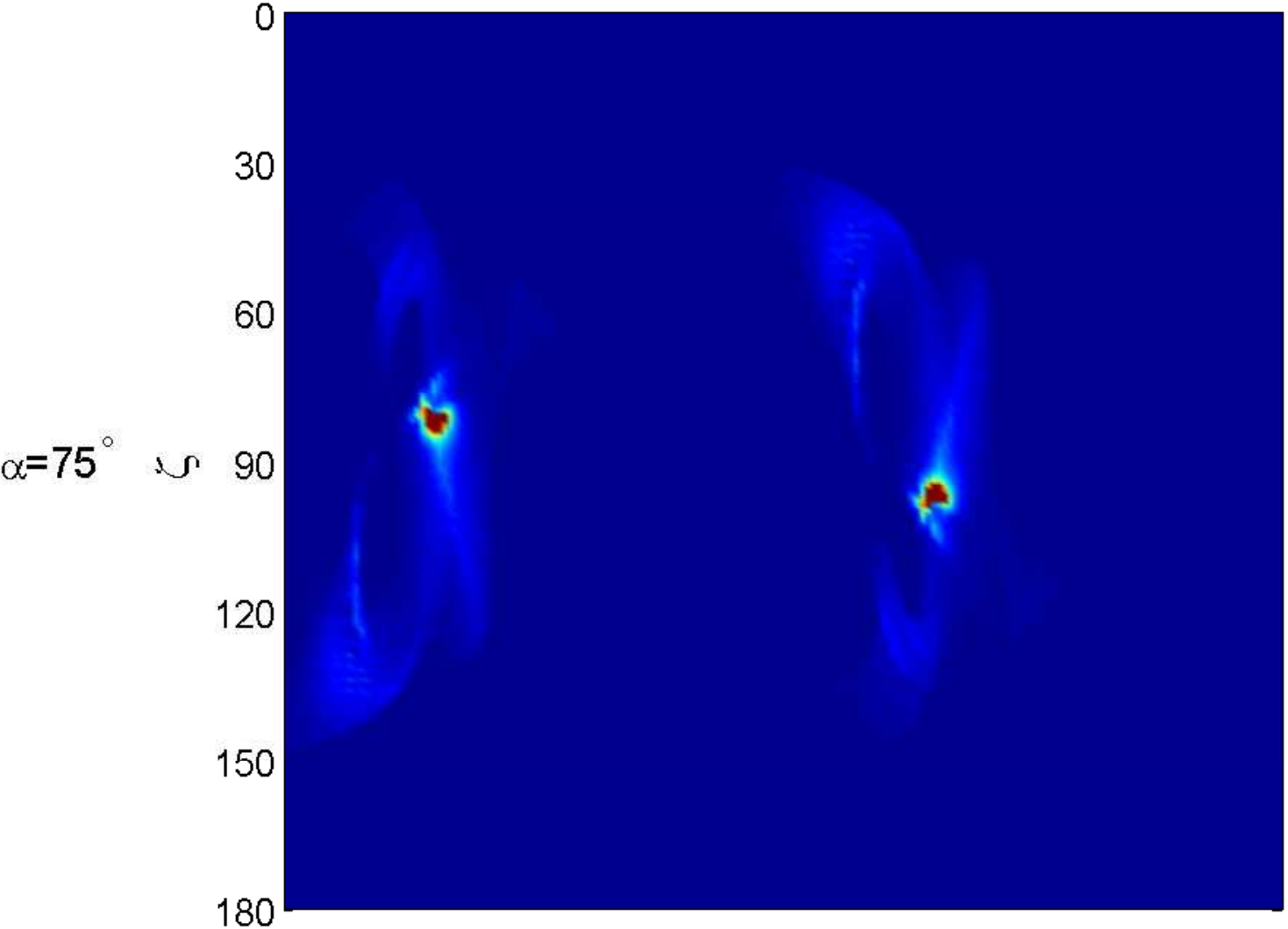} \,
\includegraphics[width=14. cm,height=3.5 cm]{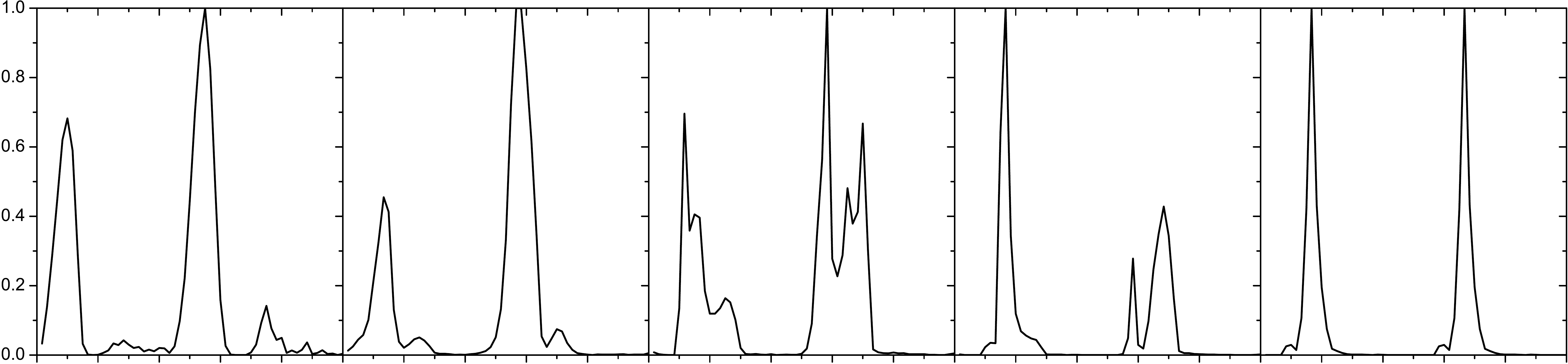} \\ \\
\includegraphics[width=4.1 cm,height=4.cm]{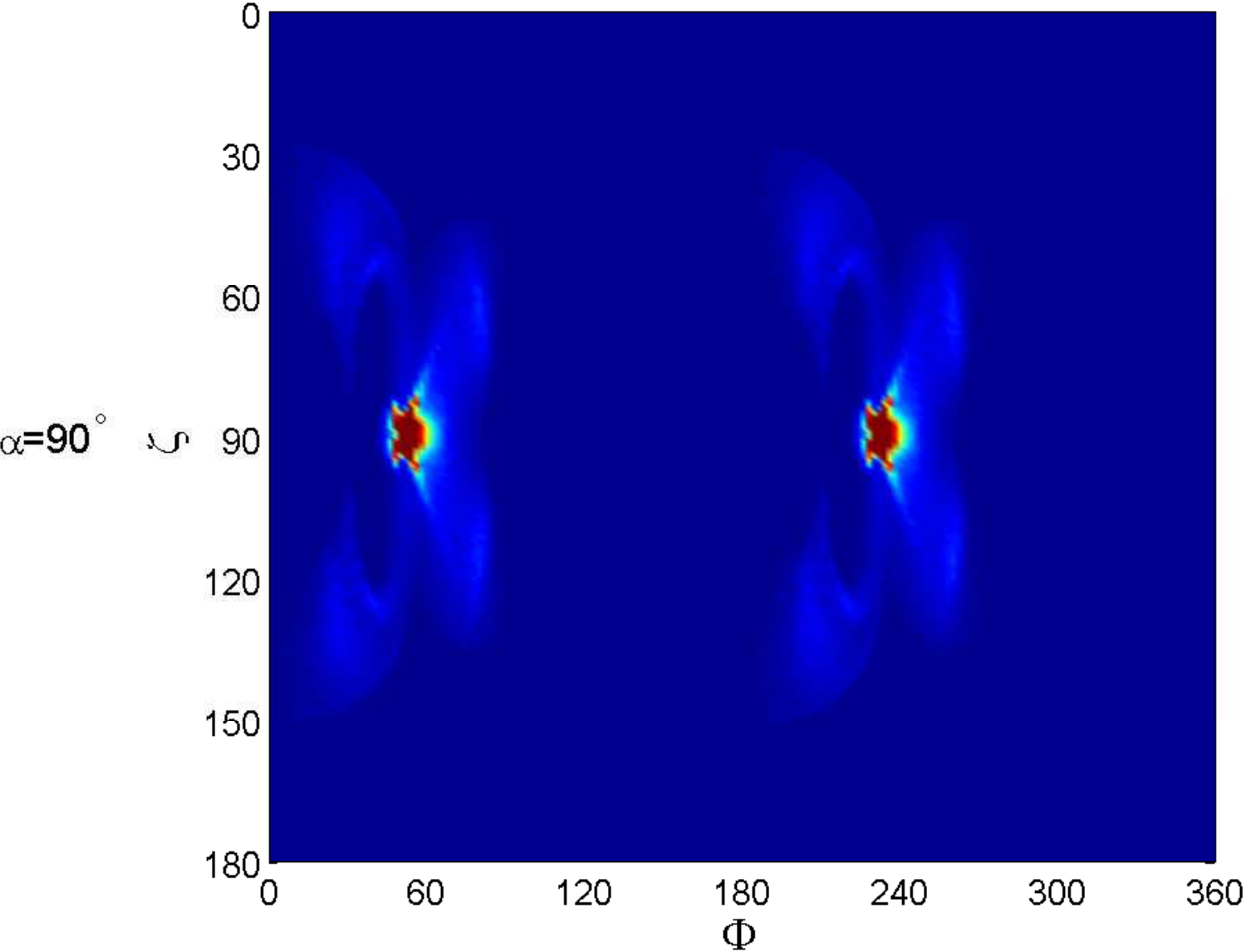} \,
\includegraphics[width=14. cm,height=4. cm]{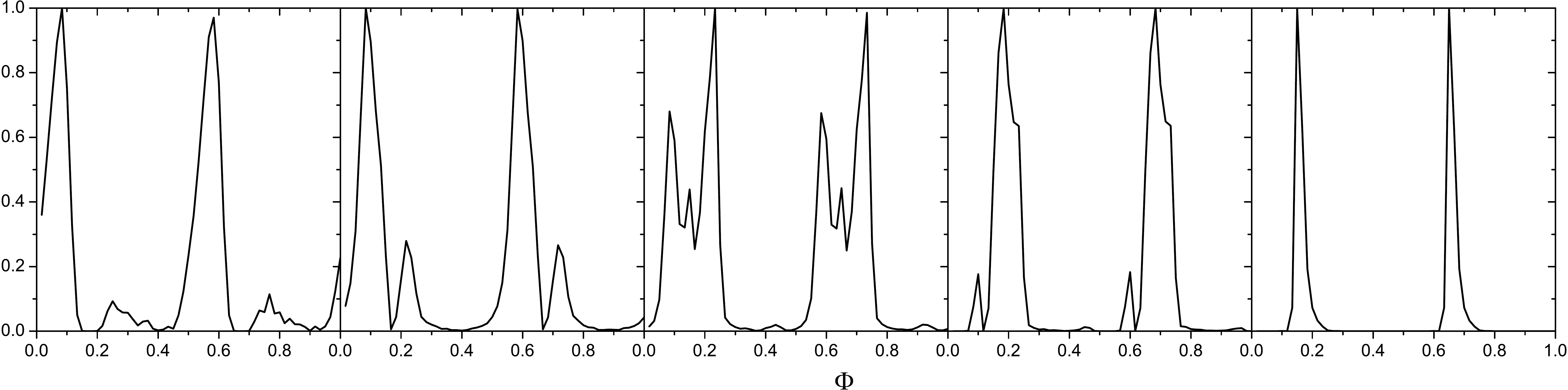} \\
\end{tabular}
\caption{Same as in Figure 7 but for the magnetosphere with a near force-free regime within the LC and with $\sigma=30 \, \Omega$ outside the LC. }
\end{figure*}
\subsection{Sky Maps and Light Curves}

The light curves can be computed by the geometry of pulsar emission models for a given magnetosphere structure. In the SG model, all accelerating electric fields are screened by the pair cascades above the pair formation fronts. There is a narrow gap with $E_{\|}\neq0$ between the last closed lines and the pair-formation fronts. The particles continue to accelerate and radiate from the neutron star surface to high altitude. The OG model is a vacuum gap that also forms  between the null charge surfaces and the last closed field lines. The outer gap extends from the null-charge surfaces to the light cylinder, and the gap width is limited by the screening of $E_{\|}\neq0$ by pair cascades. After the current through the gap is taken into account, it has been shown that the inner boundary of the outer gap is shifted inward from the null charge surface \citep{tak04,hir06}. Therefore, the gaps would exist inside the light cylinder of the pulsar magnetosphere, the details of $E_{\|}$ depend on the pair cascade process which is not taken into account in the simulation of pulsar magnetosphere so far.

To explore how the magnetic field structures and offset PCs influence $\gamma$-ray pulsar light curves, we use a geometric method to produce the $\gamma$-ray light curves based on the uniform $\sigma$ model. We assume that the $\gamma$-ray emission originates in a layer of width $\omega=0.1$ along the last closed field lines between $r_{\rm ovc}=0.9$ and $r_{\rm ovc}=1$. The emission region extends from the neutron star surface to the light cylinder, and the $\gamma$-ray emission is restricted to lie within a cylindrical radius $\rho_{\rm max}\lesssim r_{\rm L}$.
We assume the uniform emissivity along the field lines in the CF. The photon direction is assumed to be tangent to the magnetic field in the CF, obtained through a Lorentz transformation from the inertial observer's frame \citep{bai10a}. The emitted photon is collected in the sky map in viewing angle $\zeta$ and the observed phase $\Phi$, taking into account the aberration and time-delay effects. The light curves are then obtained by a cut through this sky map at constant $\zeta$.

In figure 5, we show the sky maps and the corresponding light curves from the geometric method in different inclination angles and view angles with different magnetic field configurations. We see narrow double-peak profiles at relative large $\alpha$ and $\zeta$, which are very similar to the observed $\gamma$-ray light curves.  We see that the light curves for the low $\sigma$ value are very close to that of the vacuum dipole, but the peaks are shifted in phase. As the conductivity $\sigma$ increases, the peak phase is shifted to the lager phase relative to the magnetic pole. The similar results are also found by \citet{har11} and \citet{kal12b}.
Overall, there is a significant progression in the LC shapes as $\sigma$ increases.
This is because that the magnetospheres with high conductivity have more sweepback field lines, which produces the larger shift of PC and thus causes a lager phase lag of the light curve peaks.

\section{Particle Trajectory Method}
\subsection{Real Particle Trajectories}
We study the contribution of the curvature radiation to the $\gamma$-ray light curves by including the accelerating electric fields provided by the solutions themselves.
We define the trajectory of particles in inertial observer's frame ( IOF ) by \citep{kal14}
\begin{eqnarray}
{\bf v}= \left( \frac{ {\bf E} \times {\bf B} }{B^2+E^2_{0}} + f \frac{\bf{B}}{B} \right) c ,
\end{eqnarray}
where the first term in the equation (8) is a drift velocity component, while the second term is a component parallel to the magnetic field.
By requiring that the condition $ v \simeq c$ and that the motion of the particle is outward, the spatial distribution of $f$ can be uniquely determined by equation (9). We assume that the direction of photon emission, ${\bm \eta}_{\rm em}$, is along the direction of particle motion ${ \bm \beta }={\bf v}/c$.
Therefore, the direction of photon emission in the IOF is determined by
\begin{eqnarray}
\mu_{\rm em}=\beta_{z}, \quad \phi_{\rm em}=\rm atan \left( \frac{\beta_y}{\beta_x} \right),
\end{eqnarray}
where the view angle $\zeta=\rm acos(\mu_{\rm em})$.

The observe phase is determined by include the rotation and time-delay correction
\begin{eqnarray}
\Phi=\phi_{\rm rot}-\phi_{\rm em}-{\bf r_{\rm em}} \cdot {\bm \eta}_{\rm em}/r_{\rm L},
\end{eqnarray}
where $\phi_{\rm rot}=\Omega \, dt$ is the rotation phase, $\phi_{\rm em}$ is the azimuthal angle of the emitted photon, and ${{\bf r}_{\rm em}}$ is the location of the emitted photon.

Under above assumptions, we can calculate the trajectories of radiating particles passing through each magnetospheric point in the computational domain, which also allows the determination of curvature radius $R_{\rm CR}$ along each particle trajectory. We assume that the charge particles $(e^{-} \rm,e^{+})$  are uniformly distributed on the polar cap. The $e^{\pm}$ pairs are then injected from the stellar surface with small Lorentz factor ($\gamma<100$). The Lorentz factor $\gamma$ of radiating particle along each trajectory is calculated by including the influence of the accelerating electric field and curvature radiation loss. Specifically, the $\gamma$ value along each  trajectory is integrated by the expression
\begin{eqnarray}
\frac{d\gamma}{dt}=f\frac{q_{\rm e}c E_{\|}}{m_{\rm e}c^2}- \frac{2q^2_{\rm e} \gamma^4}{3R^2_{\rm CR}m_{\rm e}c} ,
\end{eqnarray}
where $q_{\rm e}$ and $m_{\rm e}$ are the electron charge and rest mass, respectively. The first term in equation (11) is the energy gain rates of the particles due to the accelerating electric field and the second term is the energy loss rates due to curvature radiation. Then, we calculate the characteristic energy of curvature radiation $E_{\rm c}=\frac{3}{2}c\hbar\frac{\gamma^3}{R_{\rm CR}}$  and its bolometric luminosity $L_{\rm bol}=\frac{2}{3}q^2_{\rm e}c \frac{\gamma^4}{R^2_{\rm CR}}$  along each trajectory.
We integrated each particle trajectory from the neutron star surface up to $r= 2.5 \, r_{\rm L}$.
We construct the $\gamma$-ray light curve by collecting the bolometric emission with $E_{\rm c}>0.1 \, \rm GeV$ from all the emitting particles in sky maps.
For a detail description about the trajectory and radiation of particles, see \citet{kal14}.

\subsection{Sky Maps and Light Curves}


\citet{kal14} studied the $\gamma$-ray light curves using dissipative pulsar magnetospheres. It is found that the uniform $\sigma$ model cannot well match the distribution of $\gamma$-ray peak separation and the radio lag observed by Fermi-LAT. Further, they found that a significant improve can be achieved by applying a force-free regime inside the LC and a dissipative regime (FIDO) outside the LC. In fact, we also find that the uniform $\sigma$ model cannot explain the pulsar light curves observed
by Fermi LAT. Therefore, we adopt a model that is similar to the FIDO model introduced by \citet{kal14}.
\citet{kal14} used a $\sigma$ linear approximation for $E_{\|}$ based on the force-free solution at high $\sigma$ value. In this work, we adopt a more accurate treatment for $E_{\|}$. We employ a very high conductivity ($\sigma=60 \, \Omega$) within the LC and a finite conductivity outside the LC,  we then remove any $E_{\|}$ solutions within the LC. We  assume that all the $\gamma$-ray emission comes from the outer magnetosphere outside the LC. In figure 6, we shown the $E_{\|}$ distribution in the $\Omega-\mu$ plane for magnetic inclination $\alpha=60^{\circ}$ with $\sigma=60 \, \Omega$ within the LC and with $\sigma=30 \, \Omega$ outside the LC.  The field line structure is very similar to that of the force-free solution. We see the strong parallel electric field along the equatorial current sheet outside the LC. We also note that the parallel electric distribution  is very similar to that of the global PIC simulation \citep{kal18,phi18}.

In the following study, we assume the standard pulsar parameters for pulsar period $P=0.1 \, \rm s$ and  surface magnetic field $B_{\star}=10^{12} \, \rm G$. We calculate the direction of photon emission and the corresponding bolometric emission along each particle trajectory. This information allow us to produce $\gamma$-ray light curves by collecting all the photons with $E_{\rm c}>0.1 \, \rm GeV$ in sky maps. In figure 7$-$8, we show the sky maps and the corresponding light curves in different inclination angles and observer viewing angles for the magnetosphere with a near force-free regime within the LC and with  $\sigma= 0.3  \, \Omega$ and $30 \, \Omega$ outside the LC. For $\sigma=0.3 \, \Omega$ outside the LC, the light curves generally display only one broad peak. We see the double-peak light curves as $\alpha$ and $\zeta$ increase, but the peaks seem to be broad.
For very high $\sigma$ value outside the LC, we see the significant double-peak profiles except for some additional secondary peaks.
There are some cases where the light curves are narrow, which are very similar to those observed by Fermi LAT.
We find that all the emission originates in a thin layer with $r_{\rm ovc}=0.9-1$ near the equatorial current sheet outside
the LC.

\section{Application to Vela Pulsar}
The observed pulsar light curves potentially provide an important diagnostic for  pulsar magnetosphere structure. As an application, we compare the light curves predicted from the two methods with those observed at  $>0.1\,\rm GeV$ energies for Vela pulsar. The results are shown in figure 9 and 10. For the geometric method, we find that the predicted light curve with $\sigma=0.3 \, \Omega$ provides a better match to the Fermi observed data for Vela pulsar. The double-peak profile with phase separation of $\sim$ 0.43 can be better reproduced by our model with the low value of $\sigma$.
In fact, the first peak phase lag of the high $\sigma$ model is too large to explain the observed data for Vela pulsar.
We suggest that the magnetosphere with the low $\sigma$ value is more favorable for the geometric method.
For the particle trajectory method, we use the measured parameters of Vela pulsar: $P=0.089$ s, $B=3.4\times10^{12}$ G.
We find that the magnetosphere with a near force-free regime within the LC and with  $\sigma=30 \, \Omega$ outside the LC can better produce the observed data of Vela pulsar. Our results indicate that the magnetosphere with a near force-free regime within the LC and a high $\sigma$ value outside the LC is more favorable for the particle trajectory method.
\begin{figure}
\epsscale{1.}
\plotone{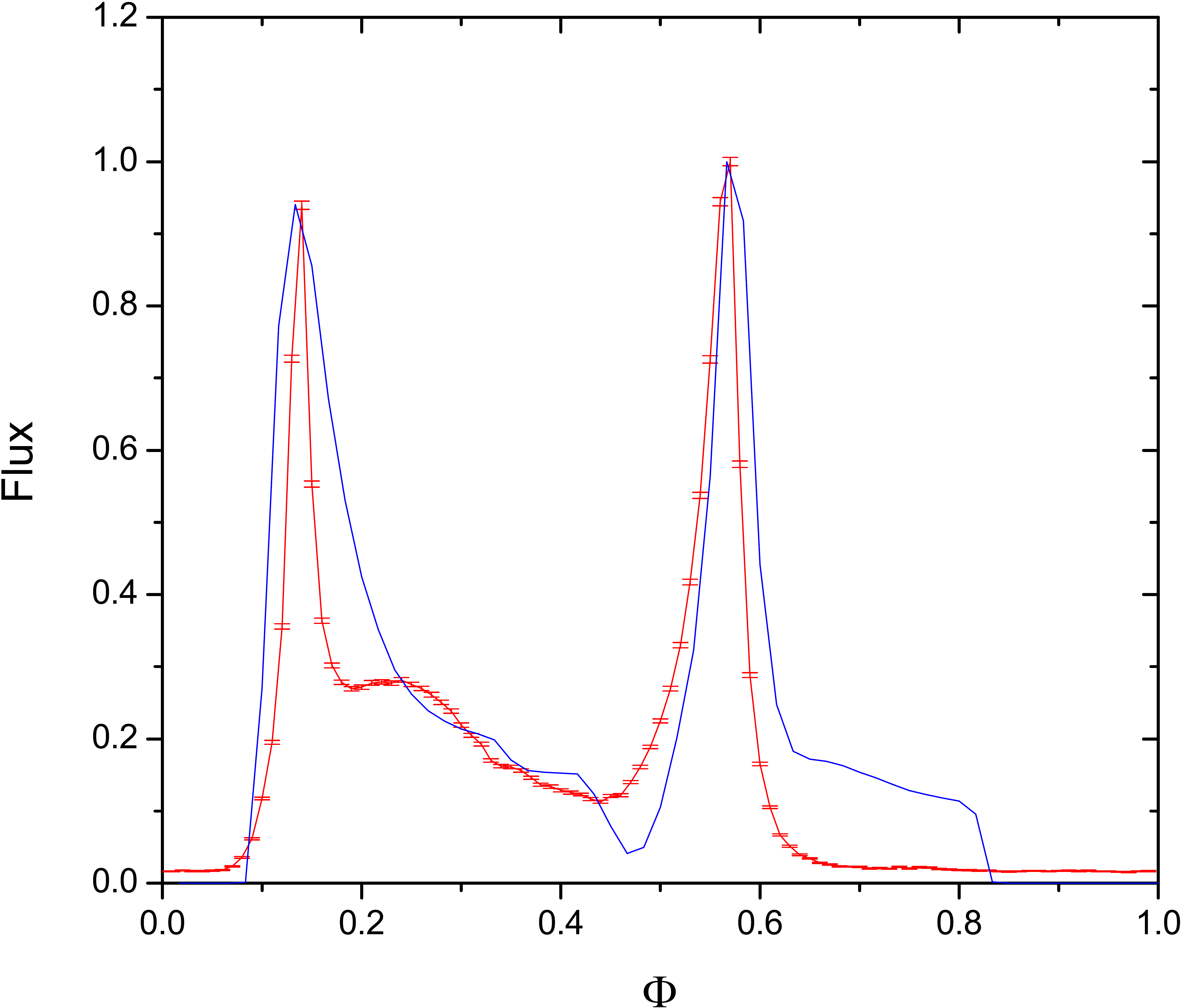}
\caption{The $\gamma$-ray light curve  at $>0.1\, \rm GeV$ energies for Vela pulsar. The red curve is the observed data taken from Abdo et al. (2013), the blue curve
is the predicted light curve from the geometric method. The model parameters are $\sigma=0.3 \, \Omega$, $\alpha=60^{\circ}$ and $\zeta=76^{\circ}$.  }
\end{figure}

\begin{figure}
\epsscale{1.}
\plotone{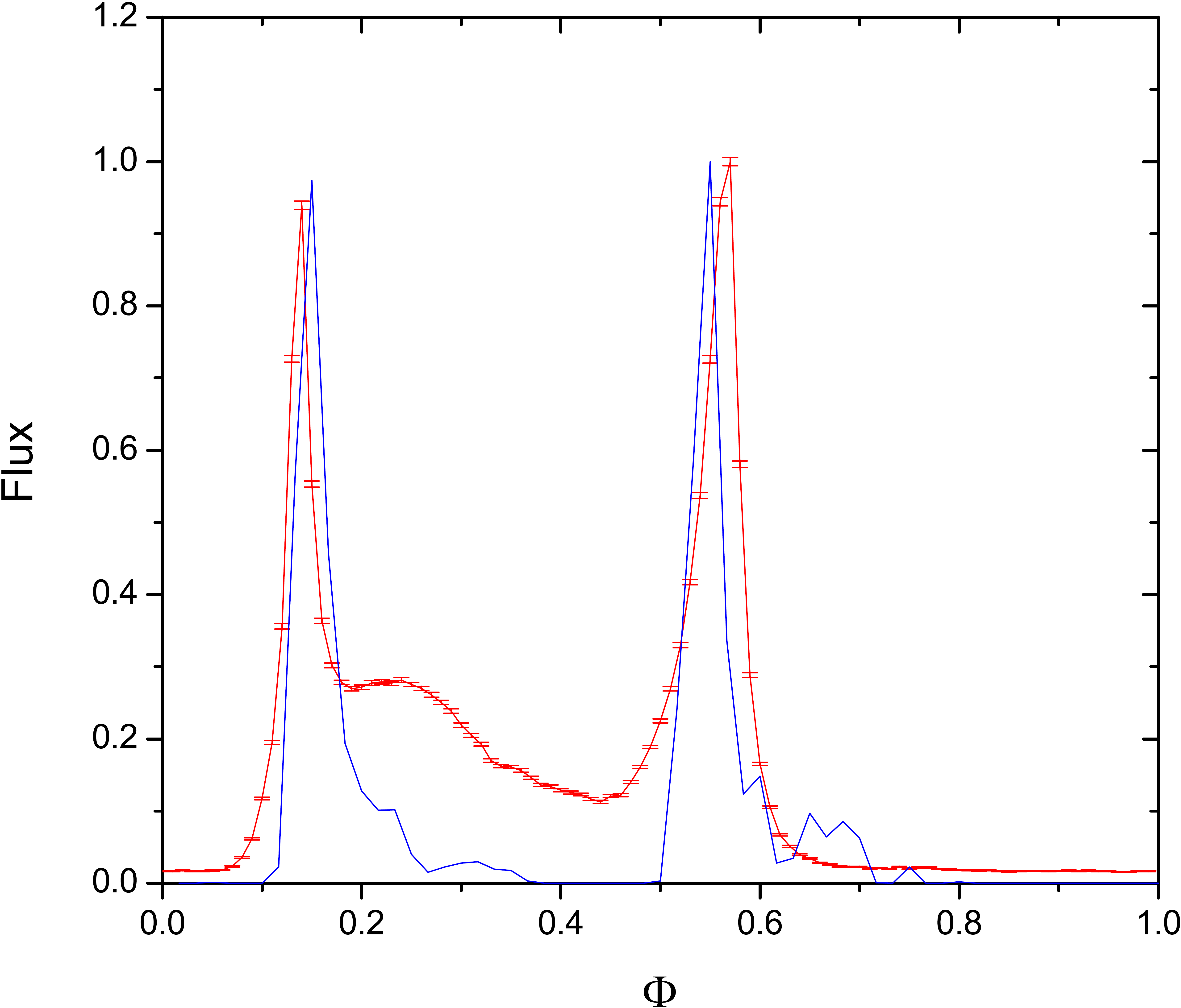}
\caption{Same as in figure 9 but for the particle trajectory method. The model parameters are $\sigma=30 \, \Omega$, $\alpha=60^{\circ}$ and $\zeta=57^{\circ}$.  }
\end{figure}
\section{Discussion and Conclusions}

In this paper, we explore the $\gamma$-ray emission patterns and light curves in dissipation pulsar magnetospheres.
The $\gamma$-ray light curves are produced by the geometric method and particle trajectory method.
For the geometric method, we assume that the $\gamma$-ray emission comes from a finite-width layer along the last closed lines, we generate the $\gamma$-ray light curves by assuming the uniform emissivity along the field lines in the CF.
We find that there is a significant progression in the $\gamma$-ray peak phase with increasing $\sigma$.
For the particle trajectory method, we consider the spatial distribution of $\sigma$ by assuming a very high conductivity within the LC and a finite conductivity outside the LC. Then, we use the field structure of these models to define realistic trajectories of radiated particles. Assuming that all the $\gamma$-ray emission comes from the outer magnetosphere outside the LC,  we compute the Lorentz factors of radiating particles and the characteristic energy of curvature radiation along each particle trajectory under the influence of both the accelerating electric field and curvature radiation loss. We produce the $\gamma$-ray  light curves by collecting the bolometric emission with $E_{\rm c}>0.1 \, \rm GeV$ from all the emitting particles. We find that the light curve shapes are very sensitive to the value of $\sigma$, and all the $\gamma$-ray emission is produced in regions near the equatorial current sheet for very high $\sigma$ value.

As an application, we compare the modeling light curves from the two methods with observed light curves at $>0.1\,\rm GeV$ energies for Vela pulsar.  For the geometric method, we find that the magnetosphere with the low $\sigma$ value provides a better match to the observed data for Vela pulsar. For the particle trajectory method, we find that the magnetosphere with a near force-free regime within the LC and a high $\sigma$ value outside the LC can better produce the light curves of Vela pulsar.
We note that the particle trajectory method uses the parallel electric fields that are self-consistent with the magnetic fields of the magnetosphere. Therefore, the results from the particle trajectory method are more physically motivated.
In fact, the microphysical processes of the pair production and particle acceleration are not self-consistently included in the simulation of the resistive magnetosphere. However, the parallel electric field distribution from the high $\sigma$ model outside the LC is only confined near the equatorial current sheet, which is very similar to that of the global PIC simulation \citep{kal18,phi18}. This gives a microphysical explanation for the origin of the parallel electric field from the high $\sigma$ model outside the LC.
Our results from the particle trajectory are also similar to those from the PIC simulation.
The particle trajectory method also allows the derivation of energy spectrum, which can be compared directly with the Fermi
observations. Therefore, we will explore the $\gamma$-ray energetic and spectral properties in future work. Further, we will improve our code by identifying  different field lines at each time-step of simulation. This will allow us apply $\sigma$ along specific magnetic field lines.  This technique will be helpful for us to understand the underlying pair production process in the gaps or current sheets.

\acknowledgments
We thank the anonymous referee for valuable comments and suggestions. We would like to thank  J\'{e}r$\hat{\rm o}$me P\'{e}tri, Gabriele Brambilla, Xuening Bai and Li Zhang for some useful discussions.  We acknowledge the financial support from the National Natural Science Foundation of China 11573060 and 11661161010, the National Science Foundation of China 11673060, the National Science Foundation of China 11871418 and Fund for Science and Technology Innovation team  in universities at Yunnan. The simulation presented in this work is performed at HPC Center, Yunnan Observatories, CAS, China.



\begin{thebibliography}{}
\bibitem[Abdo et al. (2010)]{abd10} Abdo, A. A., et al. 2010, ApJS, 187, 460
\bibitem[Abdo et al. (2013)]{abd13} Abdo, A. A., et al. 2013, ApJS, 208, 17
\bibitem[Ackermann et al. (2015)]{ack15} Ackermann, M., et al. 2015, Science, 350, 801

\bibitem[Blandford (2002)]{bla02} Blandford R. D. 2002, in Lighthouses of the Universe: The Most Luminous Celestial Objects and Their Use for Cosmology, ed. M.
Gilfanov, R. Sunyaev, \& E. Churazov (Berlin: Springer), 381
\bibitem[Belyaev (2015)]{bel15} Belyaev, M. A. 2015, MNRAS, 449, 2759
\bibitem[Bai \& Spitkovsky (2010a)]{bai10a} Bai, X.N., \& Spitkovsky, A. 2010, ApJ, 715, 1270
\bibitem[Bai \& Spitkovsky (2010b)]{bai10b} Bai, X.N., \& Spitkovsky, A. 2010, ApJ, 715, 1282
\bibitem[Brambilla et al. (2015)]{bra15} Brambilla, G., Harding, A. K., Kalapotharakos, K. \& Kazanas, D. 2015, ApJ, 804, 84
\bibitem[Brambilla et al. (2018)]{bra18} Brambilla, G., Kalapotharakos, K., Timokhin, A. N., Harding, A. K. \& Kazanas, D. 2018, ApJ, 858, 81
\bibitem[Bogovalov et al. (2018)]{bog18} Bogovalov, S. V., Contopoulos, I., Prosekin, A., Tronin, I. \& Aharonian, F. A. 2018, MNRAS, 476, 4213

\bibitem[Cao et al. (2016a)]{cao16a} Cao, G., Zhang, L., \& Sun, S. N. 2016a, MNRAS, 455, 4267.
\bibitem[Cao et al. (2016b)]{cao16b} Cao, G., Zhang, L., \& Sun, S. N. 2016b, MNRAS, 461, 1068.
\bibitem[Carrasco et al. (2018)]{car18} Carrasco, F., Palenzuela, C. \& Reula, O, 2018, Phys. Rev. D, 98, 023010
\bibitem[Chen \& Beloborodov (2014)]{che14} Chen, A. Y., \& Beloborodov A. M. 2014, ApJ, 795, L22
\bibitem[Cheng et al. (1986)]{che86} Cheng, K. S., Ho, C., \& Ruderman, M. 1986, ApJ, 300, 500
\bibitem[Cheng et al. (2000)]{che00} Cheng, K. S., Ruderman, M. \& Zhang, L. 2000, ApJ, 537, 964
\bibitem[Contopoulos et al. (1999, hereafter CKF)]{con99} Contopoulos, I., Kazanas, D., \& Fendt, C. 1999, ApJ, 511, 351
\bibitem[Cerutti et al. (2015)]{cer15} Cerutti, B., Philippov, A., Parfrey, K., \& Spitkovsky, A. 2015, MNRAS, 448, 606
\bibitem[Cerutti et al. (2016)]{cer16} Cerutti B., Philippov A. A., \& Spitkovsky A., 2016, MNRAS, 457, 2401
\bibitem[Deutsch (1955, hereafter VRD)]{deu55} Deutsch, A. J. 1955, Ann. Astrophys, 18, 1
\bibitem[Daugherty \& Harding (1982)]{dau82} Daugherty, J. K., \& Harding, A. K. 1982, ApJ, 252, 337
\bibitem[Dyks \& Rudak(2003)]{dyk03} Dyks, J., \& Rudak, B. 2003, ApJ, 598, 1201
\bibitem[Dyks et al. (2004)]{dyk04} Dyks, J., Harding, A. K., \& Rudak, B. 2004, ApJ, 606, 1125
\bibitem[Goldreich \& Julian (1969)]{gol69} Goldreich, P., \& Julian, W. H. 1969, ApJ, 157, 869
\bibitem[Etienne et al. (2016)]{eti16} Etienne, Z. B., Wan, M. B., Babiuc, M. C., McWilliams, S. T. \& Choudhary, A. 2017, Classical and Quantum Gravity, 34, 215001


\bibitem[Gruzinov (1999)]{gru99} Gruzinov, A. 1999, preprint (astro-ph/9902288)
\bibitem[Harding (2011)]{har11} Harding, A. K., DeCesar, M. E., Miller, M. C., Kalapotharakos, C., \& Contopoulos, I. 2011, arXiv:1111.0828
\bibitem[Hirotani (2006)]{hir06} Hirotani, K. 2006, ApJ, 652, 1475
\bibitem[Komissarov (2006)]{kom06} Komissarov, S. S. 2006, MNRAS, 367, 19
\bibitem[Kalapotharakos \& Contopoulos (2009)]{kal09} Kalapotharakos, C., \& Contopoulos, I. 2009, A\&A, 496, 495
\bibitem[Kalapotharakos et al. (2012a)]{kal12a} Kalapotharakos, C., Kazanas D., Harding A., \& Contopoulos, I. 2012a, ApJ, 749, 2
\bibitem[Kalapotharakos et al. (2012b)]{kal12b} Kalapotharakos, C., Harding, A. K., Kazanas, D., \& Contopoulos, I.
2012b, ApJL, 754, L1
\bibitem[Kalapotharakos et al. (2014)]{kal14} Kalapotharakos, C., Harding, A. K., \& Kazanas, D. 2014, ApJ, 793, 97
\bibitem[Kalapotharakos et al. (2018)]{kal18} Kalapotharakos, C., Brambilla, G., Timokhin, A., Harding, A. K. \& Kazanas, D. 2018, ApJ, 857, 44

\bibitem[Li et al. (2012)]{li12} Li, J., Spitkovsky, A., \& Tchekhovskoy, A. 2012, ApJ, 746, 60
\bibitem[McKinney (2006)]{mck06} McKinney, J. C. 2006, MNRAS, 368, L30
\bibitem[Muslimov \& Harding (2004)]{mus04} Muslimov, A. G., \& Harding, A. K. 2004, ApJ, 606, 1143
\bibitem[Philippov \& Spitkovsky (2014)]{phi14} Philippov, A. A. \& Spitkovsky, A., 2014, ApJ, 785, L33
\bibitem[Philippov et al. (2015)]{phi15} Philippov, A. A., Spitkovsky, A. \& Cerutti, B. 2015, ApJ, 801, L19
\bibitem[Philippov \& Spitkovsky (2018)]{phi18} Philippov, A. A. \& Spitkovsky, A., 2018, ApJ, 855, 94


\bibitem[Parfrey et al. (2012)]{par12} Parfrey K., Beloborodov A. M., \& Hui L. 2012, MNRAS, 423, 1416
\bibitem[P\'{e}tri (2012)]{pet12} P\'{e}tri, J. 2012, MNRAS, 424, 605
\bibitem[P\'{e}tri (2016)]{pet16} P\'{e}tri, J. 2016, MNRAS, 455, 3779
\bibitem[Romani \& Watters(2010)]{rom10} Romani, R. W., \& Watters, K. P. 2010, ApJ, 714, 810
\bibitem[Ruderman \& Sutherland(1975)]{rud75} Ruderman, M. A., \& Sutherland, P. 1975, ApJ, 196, 51
\bibitem[Spitkovsky (2006)]{spi06} Spitkovsky, A., 2006, ApJ, 648, L51
\bibitem[Takata et al. (2004)]{tak04}Takata, J., Shibata, S., \& Hirotani, K. 2004, MNRAS, 354, 1120
\bibitem[Tchekhovskoy et al. (2013)]{tch13} Tchekhovskoy, A., Spitkovsky, A., \& Li, J. G. 2013, MNRAS, 435, L1
\bibitem[Timokhin (2006)]{tim06} Timokhin, A. N. 2006, MNRAS, 368, 1055
\bibitem[Watters el al. (2009)]{wat09} Watters, K. P., Romani, R. W., Weltevrede, P., \& Johnston, S. 2009, ApJ, 695,
1289
\bibitem[Yu (2012)]{yu11} Yu, C. 2011, MNRAS, 411, 2461
\bibitem[Zhang \& Cheng (1997)]{zha97} Zhang, L. \& Cheng, K. S. 1997, ApJ, 487, 370
\bibitem[Zhang \& Cheng (2001)]{zha01} Zhang, L. \& Cheng, K. S. 2001, MNARS, 320, 477
\bibitem[Zhang et al. (2004)]{zha04} Zhang, L., Cheng, K. S. , Jiang, Z. J., Leung, P., 2004, ApJ, 604, 317

\end{thebibliography}
\end{document}